\documentclass[aps,prl,twocolumn,superscriptaddress,showpacs]{revtex4}

\usepackage{color}
\usepackage{graphicx}
\usepackage{ulem}
\usepackage{hyperref}
\usepackage{fancyhdr}

\definecolor{green}{rgb}{0,0.5,0}
\newcommand{\sub}[1]{_\mathrm{#1}}

\begin{document}
\pagestyle{fancy}
\lhead{Huguet J.M., Forns N., and Ritort F.}
\rhead{10 DECEMBER 2009}
\renewcommand{\headrulewidth}{1.25pt}

\title{Statistical properties of metastable intermediates in DNA unzipping}
\author{J. M. Huguet}
\affiliation{\mbox{Departament de F\'{i}sica Fonamental, Facultat de F\'{\i}sica, Universitat de Barcelona, Diagonal 647, 08028 Barcelona, Spain}}

\author{N. Forns}
\affiliation{\mbox{Departament de F\'{i}sica Fonamental, Facultat de F\'{\i}sica, Universitat de Barcelona, Diagonal 647, 08028 Barcelona, Spain}}
\affiliation{\mbox{CIBER-BBN, Networking Centre on Bioengineering, Biomaterials and Nanomedicine, ISCIII, Spain}}

\author{F. Ritort}
\email[To whom correspondence should be addressed.\\]{fritort@gmail.com}
\homepage[]{http://www.ffn.ub.es/ritort/}
\affiliation{\mbox{Departament de F\'{i}sica Fonamental, Facultat de F\'{\i}sica, Universitat de Barcelona, Diagonal 647, 08028 Barcelona, Spain}}
\affiliation{\mbox{CIBER-BBN, Networking Centre on Bioengineering, Biomaterials and Nanomedicine, ISCIII, Spain}}


\begin{abstract}
We unzip DNA molecules using optical tweezers and determine the sizes
of the cooperatively unzipping and zipping regions separating consecutive
metastable intermediates along the unzipping pathway. Sizes are found
to be distributed following a power law, ranging from one base pair up
to more than a hundred base pairs. 
We find that a large fraction of unzipping regions smaller than 10~bp are seldom detected because of the high compliance of the released single stranded DNA. We show how  the compliance of a single nucleotide sets a limit value around 0.1~N/m for the stiffness of any local force probe aiming to discriminate one base pair at a time in DNA unzipping experiments.
\end{abstract}


\maketitle
\thispagestyle{fancy}

{The mechanical response of biomolecules to externally applied forces
  allows us to investigate molecular free energy landscapes with
  unprecedented accuracy. Single molecule experiments with optical
  tweezers, atomic force microscope (AFM), and magnetic tweezers are capable of measuring forces in
  the pN range and energies as small as tenths of kcal/mol. An
  experiment that nicely illustrates the potential applications of
  single molecule manipulation is molecular unzipping
  \cite{Bockelmann:2002la,Woodside:2006kn,Danilowicz:2003ft,Rief:1999km,Koch:2002aa}. By
  applying mechanical force to the ends of biopolymers such as DNA, RNA,
  and proteins, it is possible to break the bonds that hold the native
  structure and measure free energies and kinetic rates.  In unzipping
  experiments, a DNA double helix is split into two single strands by
  pulling them apart and the force vs. distance curve (FDC) measured. A
  typical FDC shows a force plateau around 15~pN with a characteristic
  sawtooth pattern corresponding to the progressive separation of the
  two strands. Mechanical unzipping is also a
    process mimicked by motor proteins that unwind the double helix. In fact, anticorrelations between unzipping forces and unwinding rates have been found in DNA helicases suggesting that such enzymes unzip DNA by exerting local stress \cite{Johnson:2007aa}. DNA unzipping experiments have several applications such
  as identifying specific locations at which proteins and enzymes bind
  to the DNA \cite{Koch:2002aa}. Moreover, the strong dependence of the
  shape of the sawtooth pattern with the sequence might be used for DNA
  sequencing \cite{Baldazzi:2006aa}, i.e., a way to infer the DNA
  sequence from the unzipping data. A limitation factor in these
  applications is the accuracy at which base pair (bp) locations along the
  DNA can be resolved. This is mainly determined by the
    combined stiffness of the force probe and the large compliance of the released single stranded DNA (ssDNA) \cite{Bockelmann:2002la,Voulgarakis:2006aa}. The unzipping process, even if carried out
  reversibly (i.e., infinitely slowly), shows a progression of
  cooperative unzipping-zipping transitions that involve groups of bps of different sizes. These cooperatively unzipping-zipping transitions regions (CUR)
  of bps breath in an all-or-none fashion hindering details about
  the individual bps participating in such transitions. Unzipping
  experiments pose challenging questions to the experimentalist and the
  theorist. What is the typical size of these CUR? What is the smallest
  size of the CUR that can be detected with single molecule techniques?
  Under what experimental conditions might be possible to resolve
    large CUR into individual bps? There have been several
    DNA unzipping studies with controlled force using magnetic
    tweezers. Because at constant force the unzipping transition is
    abrupt, this setup is not suitable to answer such questions \cite{Danilowicz:2003ft,Lubensky:2002jb}.

We carried out DNA unzipping experiments with optical
tweezers \cite{fn:instrument, suppmat} and determined the distribution of CUR sizes in DNA
fragments a few kbp long. 
For the experiments, a 2.2 kbp DNA molecular construct was
synthesized \cite{suppmat}. In a typical unzipping experiment one bead is
held fixed at the tip of a micropipette and the other bead is optically
trapped and the force exerted on the molecule measured. By moving
the center of the optical trap at a very low speed (10~nm/s) double stranded DNA (dsDNA) is
progressively and quasireversibly converted into ssDNA through
a succession of intermediate states corresponding to the successive
opening of CUR (Fig.~\ref{fig1}a). The experimentally measured FDC shows a sawtoothlike
pattern (Fig.~\ref{fig1}b) that alternates force rips and gentle
slopes. Slopes correspond to the elastic response of the molecule
while the force rips correspond to the release of CUR.  The slope is
due to the combined elastic response of the optical trap and the
released ssDNA. The size of a CUR can be inferred from the
difference of slopes that precede and follow a given force
rip. However, the identification of the CUR sizes is not
straightforward as often the slopes cannot be isolated because the
experimental FDC exhibits noise. Here we extract the different sizes of the CUR that separate
contiguous intermediate states along the unzipping pathway. For
  that we adopt a Bayesian approach where for each experimental data
point (distance, force) we determine the most probable intermediate
state to which the data point belongs.

\begin{figure}
\includegraphics[width=8.8cm]{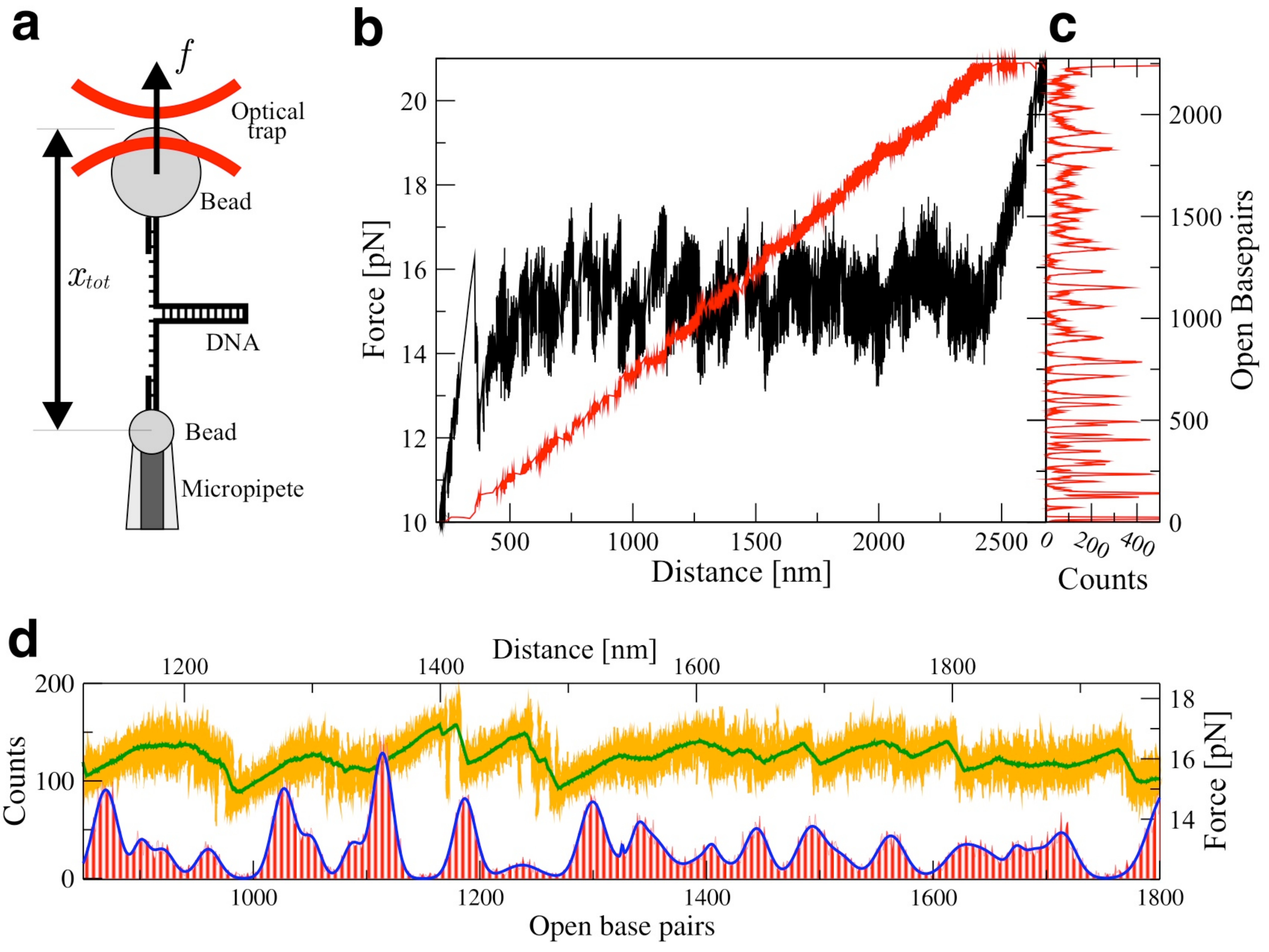}
\caption{(color online). Identifying CUR in DNA unzipping experiments. \textbf{(a)}
  Experimental setup. A 2.2 kbp sequence of DNA is unzipped using an
  optical trap and a micropipette. \textbf{(b)} Black curve shows the raw
  data of a typical FDC in an unzipping experiment. Red or gray curve shows the
  number of open bps $n^*$ corresponding to each
  experimental data point ($y$~axis of this curve is shown
  in panel c). \textbf{(c)} Histogram of the values for $n^*$
  shown in panel b. \textbf{(d)} The lower curve shows a detailed view of the histogram overlapped with a fit to a sum of Gaussians. The upper curve shows the FDC (raw data and 1~Hz low pass filtered data) corresponding to that region of the histogram.\label{fig1}}
\end{figure}

To this end we consider the molecular system as composed of different
  elements: the optical trap, the dsDNA handles, the released ssDNA and
  the hairpin at the intermediate state $I_n$ where $n$ bases are
  open. We express the total distance between trap and pipette
$x\sub{tot}$ at a given force $f$ as the sum of the extensions of
each element at that force:
\begin{equation}\label{eq:totaldistance}
x\sub{tot}(f,n)=x_{b}(f)+x_{h}(f)+x_{s}(f,n)+\frac{\phi_{b}}{2}
\end{equation}
where $x_{b}$ is the position of the bead with respect to the
center of the optical trap; $x_{h}$ is the extension of the flanking dsDNA handles;
$x_{s}$ is the extension of the released ssDNA and $\phi_{b}$ is
the diameter of the bead. The extension of the ssDNA depends on the
number of open bases at the intermediate state $I_n$. The different contributions to
  Eq.~(\ref{eq:totaldistance}) are calculated by using well-known elastic
  models for biopolymers \cite{suppmat}. For each experimental data point of the FDC
($x$, $f$),  the intermediate state $I_{n^*}$
  that passes closest to that point for a fixed force $f$ is determined by
\begin{equation}
|x-x\sub{tot}(n^*,f)|=\min_n\left(|x-x\sub{tot}(n,f)|\right).
\end{equation}
In this way each experimental data point ($x$, $f$) is associated to a
value of $n^*$ (red or gray curve in Fig.~\ref{fig1}b). The histogram
built from all values $n^*$ results in a series of sharp peaks that
can be identified with the many intermediate states $I_n$
(Fig.~\ref{fig1}c). The histogram contains information about the
stability of the intermediate states: the higher the peak, the higher
the stability of that state and the larger the GC content of that part
of the sequence (data not shown). The histogram can be fit to a sum of
Gaussians each one characterized by its mean, variance and statistical
weight (Figs.~\ref{fig1}d and \ref{fig2}a). Finally, the size of the
CUR is obtained by calculating the difference of the means (in bps) between consecutive Gaussians. The experimental distribution of
CUR sizes is shown in Fig.~\ref{fig2}b. Sizes range from a few bps up to 90 bp with a maximum number of detected CUR sizes between
20 and 50~bp. 

\begin{figure}[b]
\includegraphics[width=8.8cm]{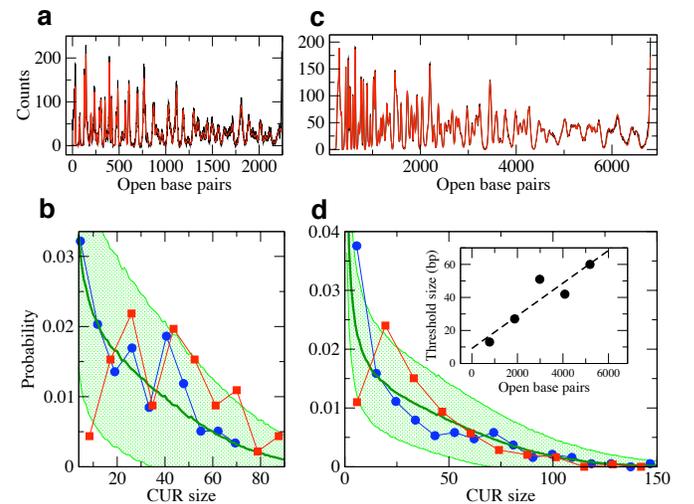}
\caption{(color online). CUR size distributions. \textbf{(a)} Histogram of $n^*$ values corresponding to
    the different
  intermediate states for the 2.2 kbp
  DNA sequence. Red or gray curve shows the experimentally measured histogram. Black curve shows
  the fit to a sum of Gaussians. \textbf{(b)} Distribution of CUR sizes for the 2.2 kbp sequence. Red or gray curve shows the experimentally mesured distribution. Blue or dark gray
  curve shows the distribution predicted by the mesoscopic model
    for DNA. Green or smooth and thick curve shows the
  distribution predicted by the toy model (see text) and the shaded
  area shows the standard deviation from different
  sequence realizations of the same length.
  \textbf{(c)} Histogram of
  intermediate states for the 6.8 kbp  sequence. Same color code as in
  panel a. \textbf{(d)} Distribution of CUR sizes for the 6.8 kbp sequence. Same color code as in panel b. \textbf{(Inset of d)} Threshold size $n^{\rm thr}$ as a
  function of the number of open bps $n$. The dashed line is a linear
  fit, $n^{\rm thr}=9.1+0.01n$. For both molecular constructs, 6 molecules have been measured and analyzed.
}\label{fig2}
\end{figure}

To better understand the distribution of CUR sizes we have computed the
sequence dependent free energy profile using a mesoscopic model for DNA based on
nearest neighbour bp interactions that includes the different elements of
the experimental setup \cite{Manosas:2005jw,Manosas:2007aa}. The model
is defined by the total free energy of the system, $G(x\sub{tot})$,
which gets contributions from the partial free energies
$G(x\sub{tot},n)$ of the many intermediates $I_n$:
$G(x\sub{tot})=-k_BT\log\{\sum_n\exp[-G(x\sub{tot},n)/k_BT]\}$. From
$G(x\sub{tot})$ we can determine the theoretical FDC by using the
relation, $f(x\sub{tot})=\frac{\partial G(x\sub{tot})}{\partial
  x\sub{tot}}$. We have used this model to compute the partial free
energies $G(x\sub{tot},n)$ of all intermediates $I_n$. For a given value
of $x\sub{tot}$ we identify the most stable intermediate $I_{n^*}$
corresponding to the value of $n^*$ for which $G(x\sub{tot},n)$ is the
absolute minimum [i.e., $G(x\sub{tot},n^*)\le G(x\sub{tot},n), \forall
n$]. Integer values of $n^*$ change in a stepwise manner as $x\sub{tot}$
is continuously varied according to the following scheme
\begin{equation}
.....\Leftrightarrow I_{n^*_a} \Leftrightarrow I_{n^*_b} \Leftrightarrow   I_{n^*_c}\Leftrightarrow....
\label{eq:reaction}
\end{equation}
where $n^*_a$, $n^*_b$, $n^*_c$ indicate the number of open bps
corresponding to consecutive intermediates. Differences between
consecutive values of $n^*$ provide the sizes of the CUR. The
resulting size distributions are shown in Fig.~\ref{fig2}b. The
good agreement between the experimental and the theoretical size
distributions shows that our method of analysis is capable of
discriminating the metastable intermediates during unzipping. There
are two remarkable facts in Fig.~\ref{fig2}b. First, the mesoscopic model
predicts a large fraction of CUR of size smaller than 10 bp that are
not experimentally observed. Second, size distributions are not smooth
but have a rough shape in agreement with the prediction by the mesoscopic
model. In order to check the generality of these
results we have repeated the same analysis by unzipping a different
and longer molecular construct of 6.8 kbp (Figs.~\ref{fig2}c,
\ref{fig2}d). The agreement between experiments and theory remains
good. Again a large fraction of predicted CUR sizes smaller than
10 bp are not detected \cite{suppmat}. However, the CUR size distributions are now
smoother suggesting that a monotonically decreasing continuous
distribution could describe the distribution of CUR in the
thermodynamic (infinite DNA length) limit. The fact that CUR sizes show a long tailed distribution indicates that large sizes
occur with finite probability. However, large-sized CUR hinder their internal
DNA sequence limiting the possibility of sequencing DNA by mechanical
unzipping. Under what experimental conditions is it possible to break up
large-sized CUR into individual bps?

In order to answer this question we have developed a toy model useful
to elucidate the mathematical form of the CUR size
distribution. Similar distributions have been investigated in the
context of DNA thermal denaturation
\cite{Ambjornsson:2006aa,Ares:2007aa} and DNA unzipping experiments in
the constant force ensemble \cite{Lubensky:2002jb}. Our model contains
only two elements: the bead in the optical trap and the DNA construct to
be unzipped. The latter is composed of the DNA duplex and the released
ssDNA (Fig.~\ref{fig3}a). The optical trap is modeled by a harmonic
spring with energy, $E_{b}(x_{b})=\frac{1}{2}kx_{b}^2$. The DNA
duplex is modeled as a one-dimensional random model with bp free
energies $\epsilon_i$ along the sequence \cite{SantaLucia:1998aa}. The
free energy of a given intermediate $I_n$ is given by
$G\sub{DNA}(n)=-\sum_{i=1}^n \epsilon_i$.  The $\epsilon_i$ are
distributed according to a normal distribution
$\mathcal{N}(\mu,\sigma)$, where $\mu (<0)$ and $\sigma$ are the mean
and the standard deviation of the energies, respectively (other more
realistic energy distributions give similar results). The released
ssDNA is taken as inextensible: its extension ($x_{m}$) is given by
$x_{m}=2dn$, where $d$ is the interphosphate distance, $n$ is the
number of open bps, and the factor 2 stands for the two strands of
ssDNA.  By using the relation $x_{b}=x\sub{tot}-2dn$
(Fig.~\ref{fig3}a), the total energy of the system can be written as
\begin{equation}\label{eq:etot}
E(x\sub{tot},n)=\frac{1}{2}k(x\sub{tot}-2dn)^2-\sum_i^n \epsilon_i.
\end{equation}
At fixed $x\sub{tot}$, the system will occupy the state ($n^*$) that minimizes the total energy of the
system, i.e., $E(x\sub{tot},n^*)\le  E(x\sub{tot},n), \forall
  n$. The function $n^*(x\sub{tot})$ gives the thermodynamic energy
  function at the minimum, $E_m(x\sub{tot})$, and the FDC, $f(x_{tot})=\frac{\partial E_{m}(x\sub{tot})}{\partial x\sub{tot}}$. The
FDC obtained from this model reproduces the sawtooth pattern that is
experimentally observed (Fig.~\ref{fig3}b).

\begin{figure}[t]
\includegraphics[width=8.8cm]{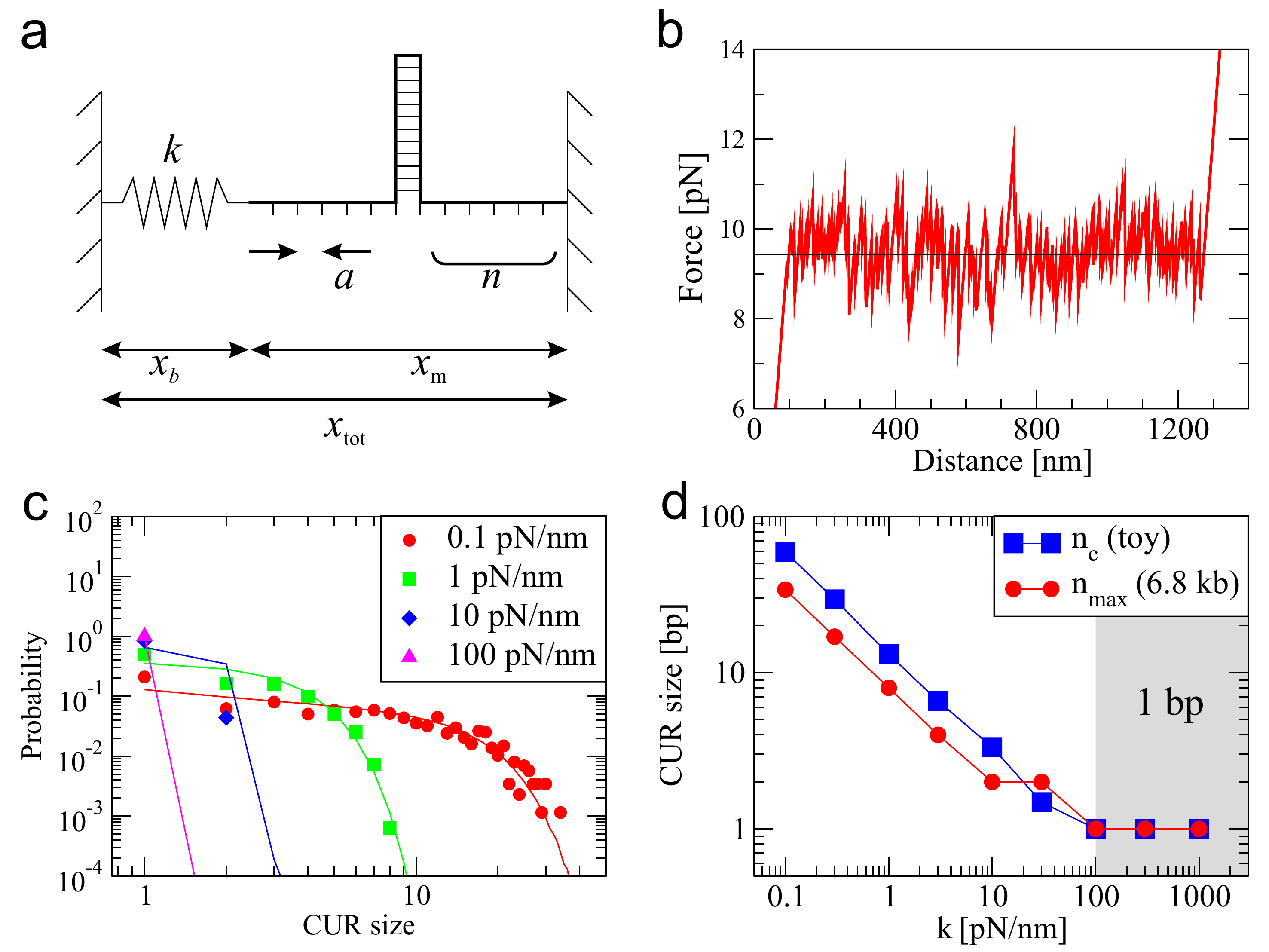}
\caption{(color online) Toy model. \textbf{(a)} The unzipping experiment is
  modeled with the minimum elements of the experimental
  setup. \textbf{(b)} Red or gray curve shows the FDC for one random
  realization. The horizontal black line shows the analytical
  approximation when disorder is neglected. \textbf{(c)} CUR size
  distributions in log-log scale for some values of $k$ using the toy
  model. Data plotted with points shows the CUR size distribution for
  the 6.8~kbp sequence. Data plotted with lines, shows the average CUR
  size distribution over $10^4$ realizations ($k$=60~pN/$\mu$m,
  $d=0.59$~nm, $\mu$=-1.6~kcal/mol,
  $\sigma$=0.5~kcal/mol). \textbf{(d)} The fit of the average CUR size
  distributions in panel c to Eq.~\ref{eq:fit} give the cutoff size
  $n_c$. It decreases like $n_c\simeq k^{-2/3}$. Blue or dark gray curve shows $n_c$ vs~$k$. Red or gray curve shows the maximum
  CUR size ($n_{max}$) predicted by the toy model for the 6.8~kbp
  sequence. For $k>100$~pN/nm, both curves level off to CUR sizes of 1~bp. \label{fig3}}
\end{figure}

Equation \ref{eq:etot} can be approximated by neglecting the disorder and taking
$\epsilon_i=\mu,\forall i$. This gives, 
\begin{equation}\label{eq:etotmean}
E(x\sub{tot},n)\simeq\frac{1}{2}k(x\sub{tot}-2dn)^2-\mu n.
\end{equation}
From this approximation we immediately get the following results
$n^*(x\sub{tot})=\frac{1}{2d}\left(x\sub{tot}+\frac{\mu}{2dk}\right)$,  
$E_m(x\sub{tot})=-\frac{\mu}{2d}\left(x\sub{tot}+\frac{\mu}{4kd}\right)$, and $f\sub{m}=-\frac{\mu}{2d}$. These expressions capture the
dependence of the averaged number of open bps, energy, and force
on the external parameters $\mu,\sigma$ \cite{suppmat}.

Finally we have numerically computed the CUR size
  distribution. We find that this mostly depends
on $\sigma$ and $k$.  For several combinations of $\sigma$ and $k$ we simulated $10^4$ realizations (i.e., sequences) of
$10^4$ bp sequences, while $d$ and $\mu$ were
kept constant. The size distributions are excellently fit by
  a power law with a superexponential cutoff \cite{suppmat}:
\begin{equation}\label{eq:fit}
P(n)=An^{-B}\exp\left(-(n/n_c)^C\right),
\end{equation}
where $P(n)$ is the probability of observing a CUR of size $n$;
$A,B,C$ and $n_c$ (cutoff size) are positive fitting parameters. How
much can the toy model predict the experimental results? For the 2.2~kbp sequence the parameters that best fit the experimental histograms
are $\mu=-2.80$~kcal/mol, $\sigma=2.2$~kcal/mol and $k=60$~pN/$\mu$m
(equal to the stiffness of the trap that we can measure
independently). This gives $A=0.058, B=0.42, C=2.95$, and $n_c=69$ (fit
shown in Fig.~\ref{fig2}b). For the 6.8~kbp sequence we find
$\sigma=3.3$~kcal/mol while $k$ and $\mu$ have the same value. This
gives $A=0.050, B=0.43, C=3.0$, and $n_c=91$ (fit shown in
Fig.~\ref{fig2}d). The values of $\mu$ and $\sigma$ are not far from
the actual mean and standard deviation of the energies of the nearest
neighbour model for DNA, $\mu$=-1.6~kcal/mol, $\sigma=0.44$~kcal/mol
\cite{SantaLucia:1998aa}. Having not included the elastic effects of
the ssDNA in the toy model we should not expect a good match between
the fitting and the experimental values.

What is the limiting factor in detecting small-sized CUR? A look at
Figs.~\ref{fig1}c,\ref{fig2}a,\ref{fig2}c}, and Figs.~S14 and S15 in
\cite{suppmat} shows that histograms become smoother as the molecule is
progressively unzipped. The increased compliance of the molecular setup
as ssDNA is released markedly decreases the resolution in discriminating
intermediates. In fact, for the 6.8 kbp construct we found that along the
first 1500~bp of the hairpin only 30\% of the total number CUR smaller
than 10~bp are detected whereas beyond that limit no CUR smaller than
that size is discriminated. If we define the threshold size $n^{\rm
  thr}$ as the size of the CUR above which $50\%$ of the predicted CUR
are experimentally detected we find that $n^{\rm thr}$ increases
linearly with the number of open bps putting a limit around 10~bp
for the smallest CUR size that we can detect (Fig.~\ref{fig2}d,
inset). What is the limiting factor in resolving large-sized CUR
  into single bps? Only by applying local force on the opening
  fork (thereby avoiding the large compliance of the molecular setup)
  and by increasing the stiffness of the probe might be possible to
  shrink CUR size distributions down to a single bp \cite{Voulgarakis:2006aa}. Figures~\ref{fig3}c and \ref{fig3}d show how the CUR size
  distributions shrink and the largest CUR size decreases as the
  stiffness increases. Its value should be around 50-100~pN/nm for all
  CUR sizes to collapse into a single bp. Remarkably enough
  this number is close to the stiffness value expected for an individual
  DNA nucleotide strecthed at the unzipping force \cite{suppmat}. Any probe more rigid
  than that will not do better. Similarly to the problem of atomic
  friction between AFM tips and surfaces we can define a parameter $\eta$ (defined as the
  ratio between the rigidities of substrate and cantilever) that controls the
  transition from stick slip to continuous motion \cite{Socoliuc:2004aa}. For DNA unzipping we have
  $\eta=\frac{|\mu|}{k d^2}$ where $\mu$ is the average free energy of
  formation of a single bp, $k$ is the probe stiffness, and $d$ is the interphosphate distance. The value $\eta=1$ determines the
  boundary where all CUR are of size equal to one bp ($\eta<1$). In our experiments we have $\eta \simeq 500$ and to
  reach the boundary limit $\eta= 1$
  we should have $k\sim$ 100 pN/nm consistently with what is shown in
  Figs~\ref{fig3}c and \ref{fig3}d. It is remarkable that the elastic
  properties of ssDNA lie just at the boundary to allow for one bp
  discrimination. This suggests that molecular motors that mechanically
  unwind DNA can locally access the genetic information one bp at a  time \cite{suppmat}. 

In summary, we have measured the distribution of sizes of unzipping regions of DNA. A toy model reproduces the experimental results and can be used to infer the experimental conditions under which the unzipping is done one bp at a time. This is achieved when the stiffness of the probe is higher than 100~pN/nm, which coincides with the stiffness of one base of ssDNA at the unzipping force.

J.M.H. acknowledges a grant from the Ministerio de
Educaci\'on y Ciencia in Spain. F.R. is supported by grants No.~FIS2007-3454 and No.~HFSP (RGP55-2008).




\end{document}


\title{\small{Supplementary information of the paper}\\\Large{Statistical properties of metastable intermediates in DNA unzipping}}

\author{J. M. Huguet}
\affiliation{Departament de F\'{\i}sica Fonamental, Facultat de F\'{\i}sica, Universitat de Barcelona, Diagonal 647, E-08028, Barcelona}

\author{N. Forns}
\affiliation{Departament de F\'{\i}sica Fonamental, Facultat de F\'{\i}sica, Universitat de Barcelona, Diagonal 647, E-08028, Barcelona}
\affiliation{CIBER de Bioingenier\'{\i}a, Biomateriales y Nanomedicina, Instituto de Salud Carlos III, Madrid}

\author{F. Ritort}
\email[To whom correspondence should be addressed: ]{fritort@gmail.com}
\homepage[]{http://www.ffn.ub.es/ritort/}
\affiliation{Departament de F\'{\i}sica Fonamental, Facultat de F\'{\i}sica, Universitat de Barcelona, Diagonal 647, E-08028, Barcelona}
\affiliation{CIBER de Bioingenier\'{\i}a, Biomateriales y Nanomedicina, Instituto de Salud Carlos III, Madrid}

\date{10 December 2009}

\maketitle
\tableofcontents
\vspace{2cm}

\def\thesection{\arabic{section}} 
\def\thesubsection{\arabic{section}.\arabic{subsection}} 
\def\thesubsubsection{\arabic{section}.\arabic{subsection}.\arabic{subsubsection}} 
\def\thefigure{S\arabic{figure}} 

\section{Experimental details}
\subsection{DNA sequences}
\begin{figure}[!h]
\includegraphics[width=9cm]{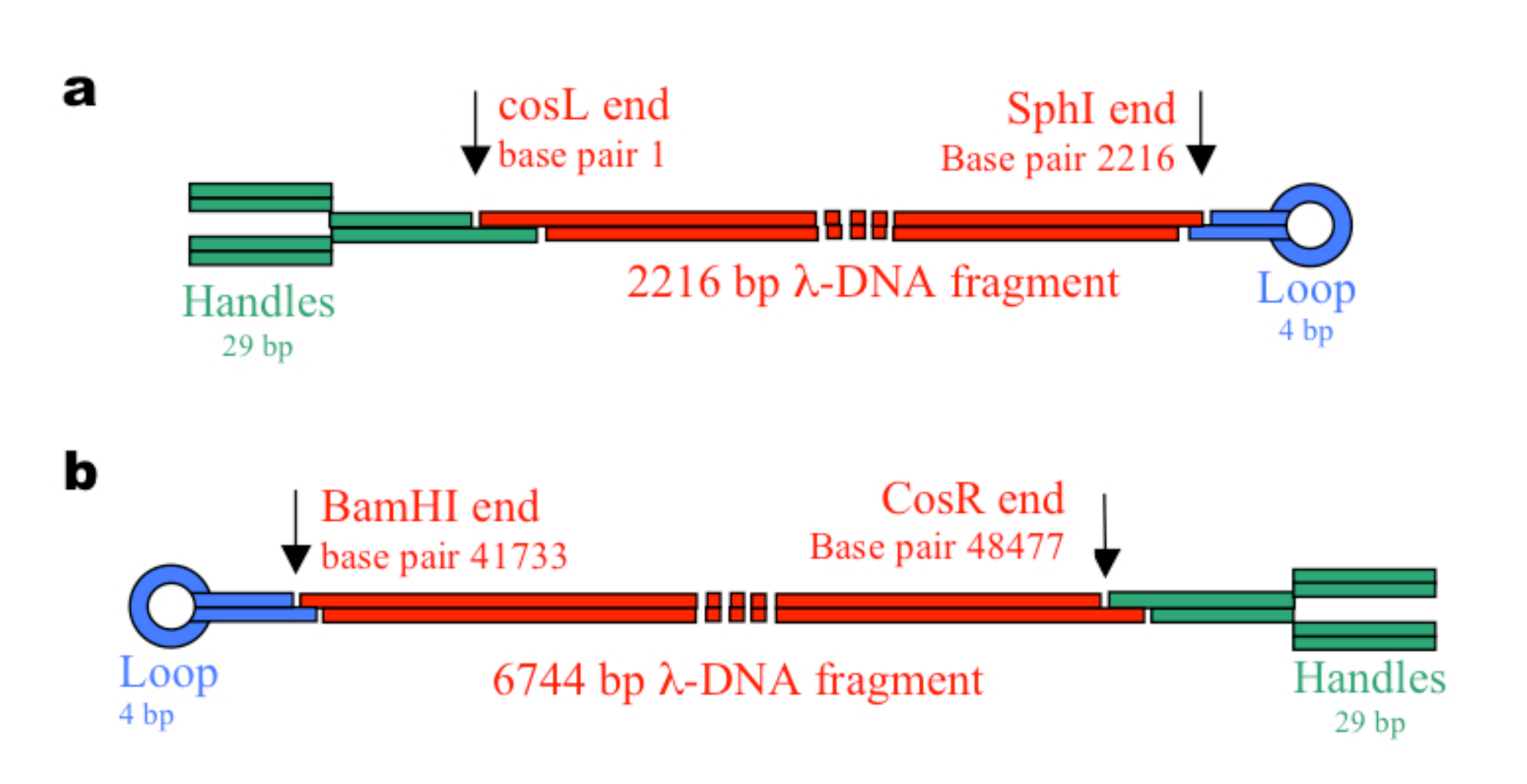}
\caption{DNA molecules unzipped. \textbf{a.} The 2.2 kb sequence. The main sequence (in red) corresponds to the bases from 1 to 2216 of the $\lambda$ genome. The handles are annealed at the cosL end and the molecule is unzipped from 5' to 3' direction. The linking fragment (duplex of DNA in green) between the handles and the main sequence  is 5'-aatagagacacatatataatagatctt-3'. The linking fragment (duplex of DNA in blue) between the main sequence and the loop is 5' tgatagcct-3'. \textbf{b.} The 6.8 kb sequence. The main sequence (in red) corresponds to the bases from 41733 to 48477 of the $\lambda$ genome. The handles are annealed at the cosR end and the molecule is unzipped along the 3' to 5' direction. The fragment between the main sequence and the handles is 5'-gggcggcgacctaagatctattatatatgtgtctctatt-3'. The fragment between the loop and the main sequence is 5'-aatagagacacatatataatagatctt-3'.}\label{fig:sequences}
\end{figure}

The two DNAs (2.2 and 6.8 kb) unzipped in the experiments are obtained by gel extraction from $\lambda$-DNA. They are synthesized following similar procedures: 1) A restriction enzyme is used to cleave the $\lambda$-DNA molecule. The SphI restriction enzyme (BamHI) is used for the 2.2 kb (6.8 kb) sequence. 2) A fragment of 2216 bp (6770 bp) is isolated from the resulting digestion by gel extraction. 3) Two handles of 29 bp are added by hybridization and annealing of short complementary oligonucleotides to one end of the 2.2 kb (6.8 kb) fragment. Another oligonucleotide that forms a tetraloop (5'-acta-3') is also annealed at the other end of the 2.2 kb (6.8 kb) fragment. The handles were labeled with biotin and digoxigenin that specifically attach to coated polystyrene beads. Figure \ref{fig:sequences} shows the resulting sequences. Experiments were done in aqueous
  buffer containing 10 mM Tris-HCl (pH 7.5), 1 mM EDTA, 500 mM NaCl and
  0.01\% Sodium Azide. 

\subsection{Measurements calibration and data acquisition}
The instrument has a force resolution below 1 pN, which represents about ~6\% uncertainty at the mean unzipping force (16.5 pN). The force is inferred by measuring the deflection of the scattered light by the bead. The offset of the deflected light is measured using a Position Sensitive Detector (PSD), which is converted into force by multiplying it with a calibration factor. The force is calibrated using three different methods and all agree within the aforementioned uncertainty: power spectrum measurements, the Stokes law and the equipartition theorem. The distance measurements have a resolution of 1 nm which represents about ~3\%. The distance is measured with a light-lever. A small amount of the laser beam is split before it enters the focusing objective and forms the optical trap. The light is redirected to a Position Sensitive Detector (PSD) that measures the position of the center of the optical trap. The PSD is calibrated using a motorized stage with known pitch distance. 

The analog signals from the PSDs (position and force) are filtered using an analog low pass filter of bandwidth 1 kHz. The resulting signal is sampled at 4 kHz producing the raw data that we obtain in the unzipping experiment.

\subsection{Statistics and reproducibility of measurements}
Six different molecules were analyzed for the 2.2 kb and the 6.8 kb DNA sequences. 

\begin{figure}[!h]
\includegraphics[width=10cm]{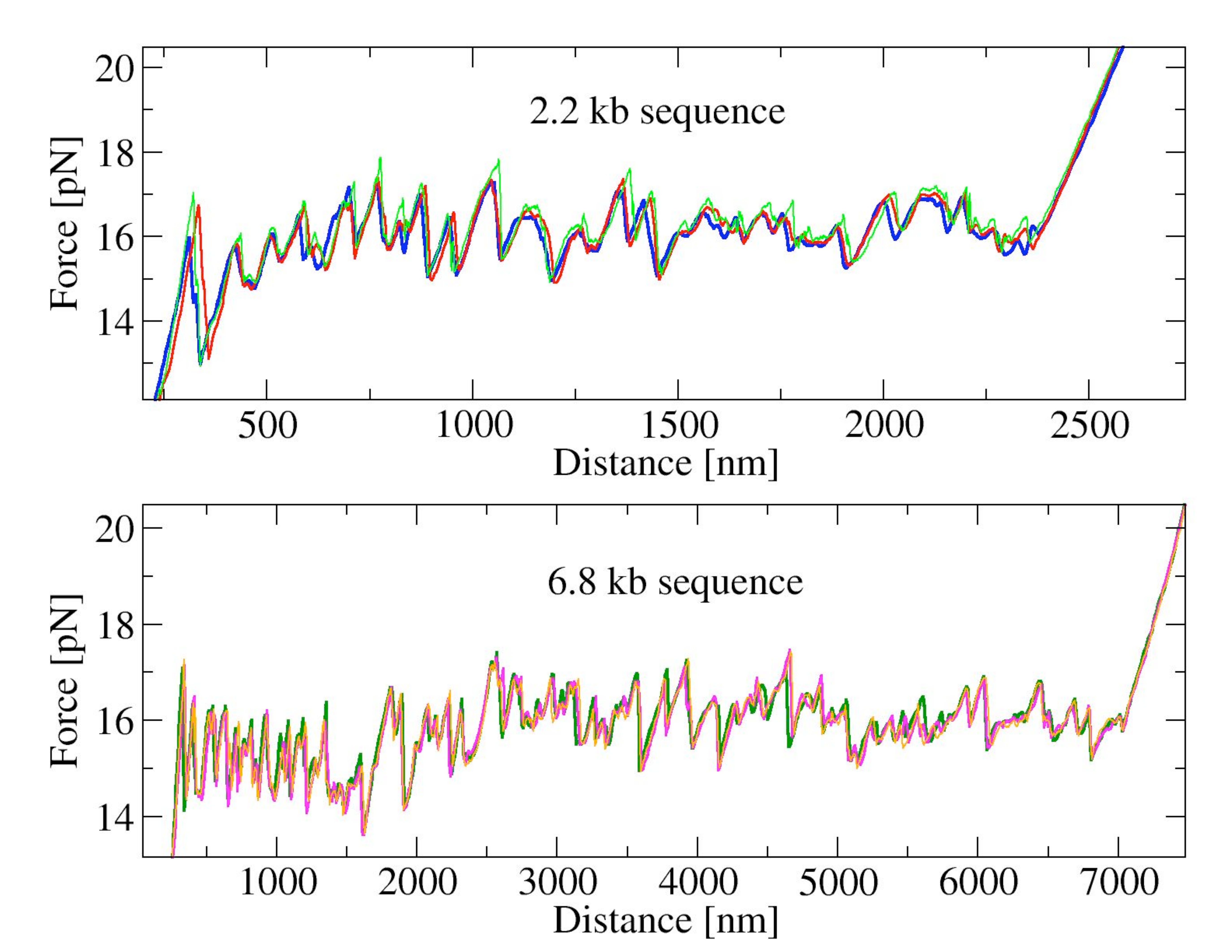}
\caption{Experimental FDC of 3 molecules corresponding to the 2.2 kb sequence (upper panel) and the 6.8 kb sequence (lower panel). Since the raw data is too noisy, the data has been filtered with a low-pass running-average filter with a bandwidth of 1 Hz to clearly see the traces.}\label{fig:overlaying}
\end{figure}

In fig \ref{fig:overlaying} we show force-distance curves measured for 3 different molecules corresponding to the 2.2 kb and 6.8 kb sequences. As can be seen our measurements are reproducible. Slight differences between different traces are due to the variability of the molecular setup and instrumental drift effects. 

\section{Experimental errors}

\subsection{Parameters used in the model}\label{sec:parameters}
\begin{figure}[!h]
\includegraphics[width=10cm]{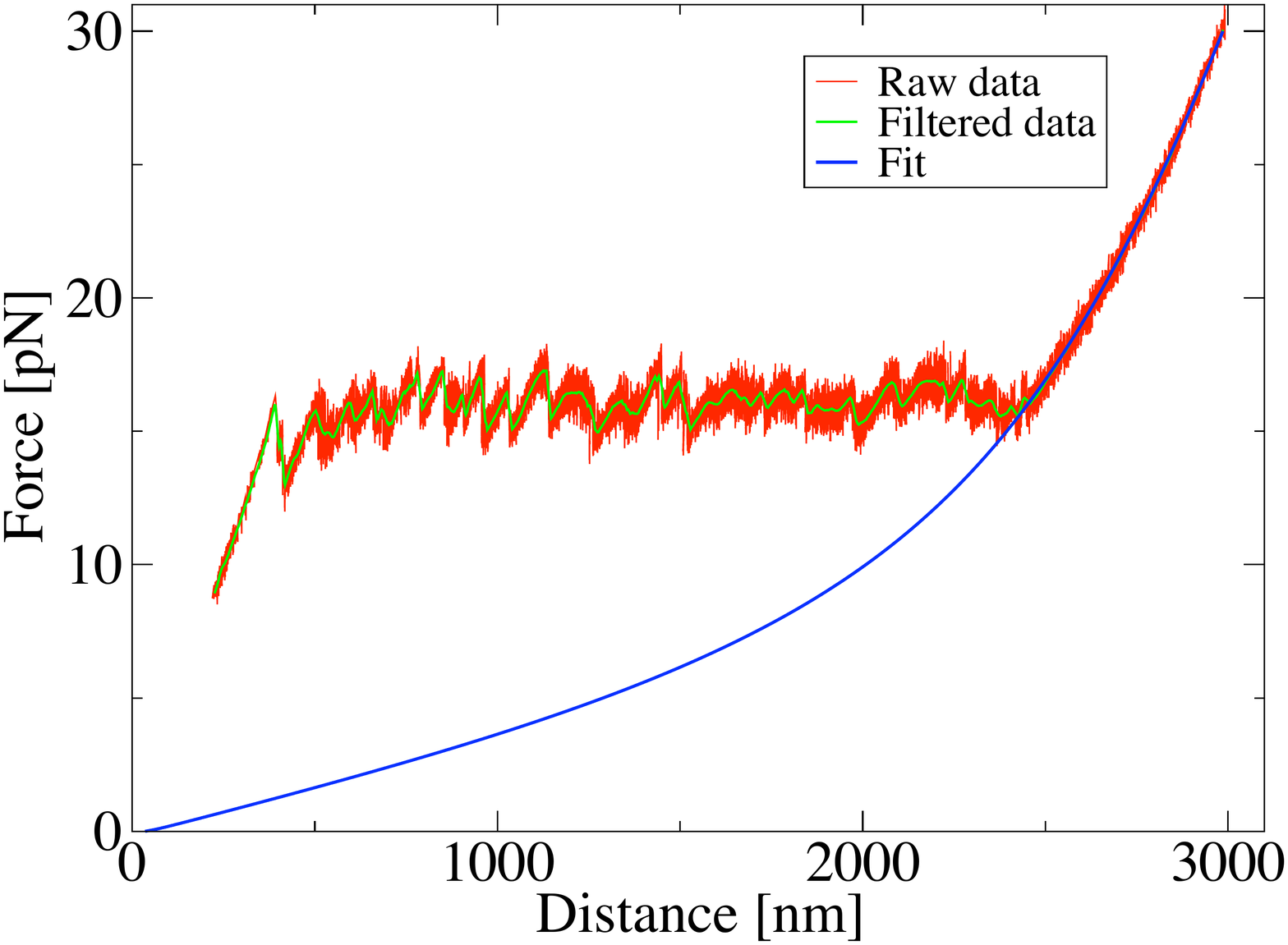}
\caption{Fit of the Kuhn length of the ssDNA using the last part of the FDC to one 2.2 kb molecule.}\label{fig:parameterfit}
\end{figure}

Here we give the numerical values of the parameters used in the model (Eq.~(1) main text) to extract the histogram of intermediate states. The bead in the optical trap is modeled by a Hookean spring, $f=kx_b$, where $k$ is the measured stiffness of the optical trap and $x_b$ is the position of the bead with respect to the center of the optical trap. The measured stiffness of the optical trap is $k=60\pm5$~pN/nm (error is due to bead size heterogeneity). The elastic response of the handles is described by the worm-like chain (WLC) model and parameters are obtained from the literature \cite{Bustamante:1994aa}. The elastic response of the released ssDNA is described using a freely-jointed chain (FJC) model \cite{Smith:1996wc}. We use $d=0.59$ nm for the ssDNA because this value fits well the elastic response of the ssDNA in our data. This value is similar to the one found by Dessinges et al. \cite{Dessinges:2002aa} ($d$=0.57 nm at 1 and 10 mM phosphate buffer) and to Johnson et al. \cite{Johnson:2007aa} ($d$=0.537 nm at 50~mM NaCl, 20 mM Tris-HCl, pH 7.5). Moreover we consider that interphosphate distance is an effective parameter of an elastic model which does not need to be equal to the parameter measured by crystallography. In order to determine the Kuhn length of the ssDNA, we have proceeded as follows (see fig \ref{fig:parameterfit}): We fix the interphosphate distance at $d$=0.59 nm and then we determine the Kuhn length of the ssDNA by fitting the last part of the FDC (where the dsDNA duplex is fully unzipped and the elastic response of the ssDNA can be measured) to a Freely Jointed Chain. The best value among the 12~molecules (6~molecules of each) for the Kuhn length is $b=1.2\pm0.3$~nm. We assume an inextensible Freely Jointed Chain model because the effect of a stretch modulus on the elastic response of the ssDNA is barely negligible below 20 pN, where the unzipping is observed.

\subsection{About using force-distance curves (FDCs) instead of force-extension curves (FECs)}
\begin{figure}[!h]
\includegraphics[width=10cm]{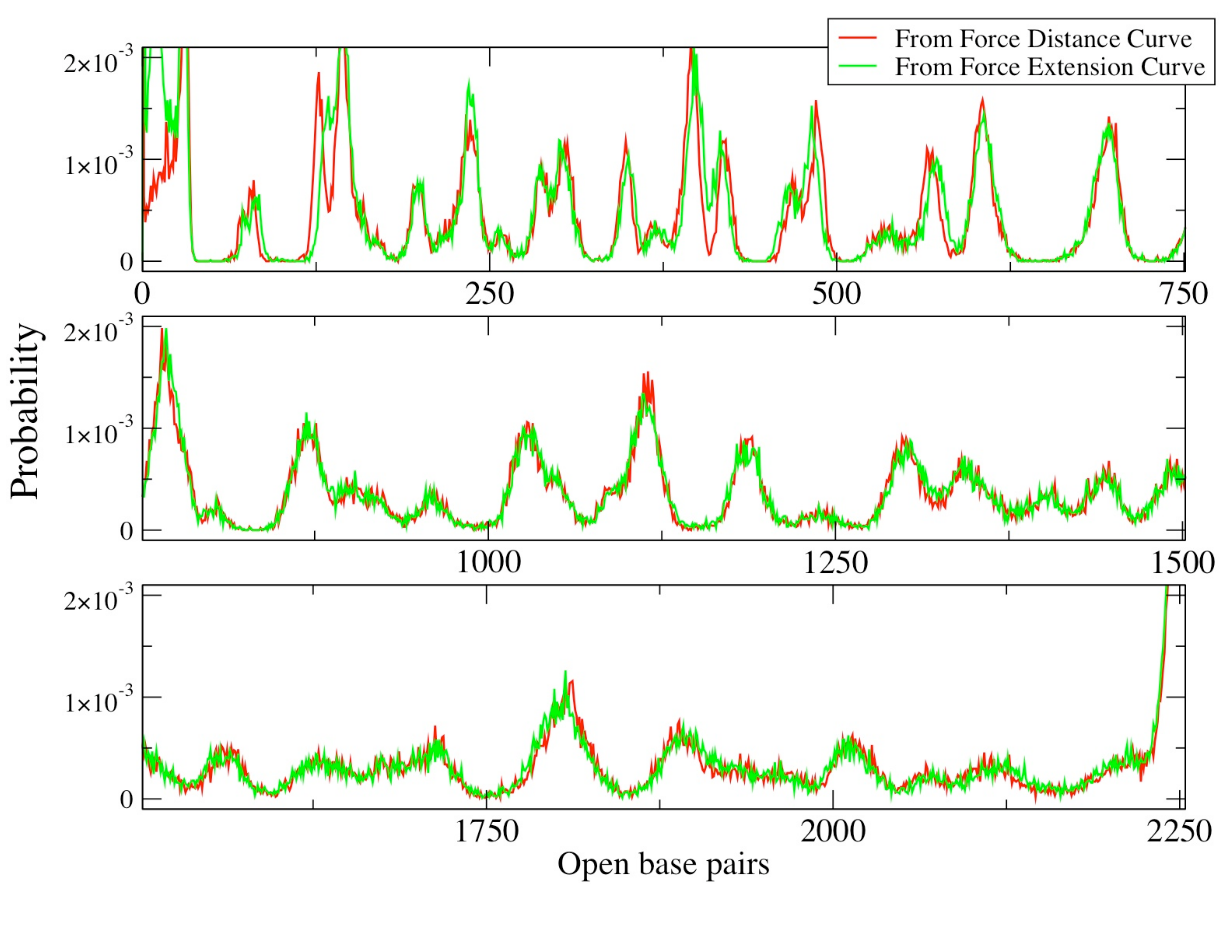}
\caption{Histograms of intermediate states calculated using the FDC and the FEC. Note that, apart from some peaks in the beginning, there are no significant differences in the position of the peaks. Eventually, the distribution of CUR sizes will hardly be affected by the choice of FDC or FEC.}\label{fig:histFDEC}
\end{figure}

We define the distance $x_{tot}$ as the length between the bead of the micropipette and the center of the optical trap (see Fig.~1a for an illustration of how $x_{tot}$ is defined and Eq.~(1) for a mathematical expression). This magnitude is a measurement that we collect directly from the instrument as the optical trap is moved up and down along the fluidics chamber. This is the control parameter in the experiment, i.e. the variable that does not fluctuate and the parameter that determines the statistical ensemble (what we call mixed ensemble). Since we know the trap stiffness ($k$) and we measure the total distance ($x_{tot}$) and the force ($f$), it is straightforward to convert the force-distance curve into a force-extension curve using the following relation: $x_m=x_{tot}-f/x_b$, where $x_m$ is the molecular extension. As we show in the figure \ref{fig:histFDEC} we do not appreciate any significant difference when computing the histogram using a FEC or a FDC. 

\subsection{Base pair determination}
\begin{figure}[!h]
\includegraphics[width=10cm]{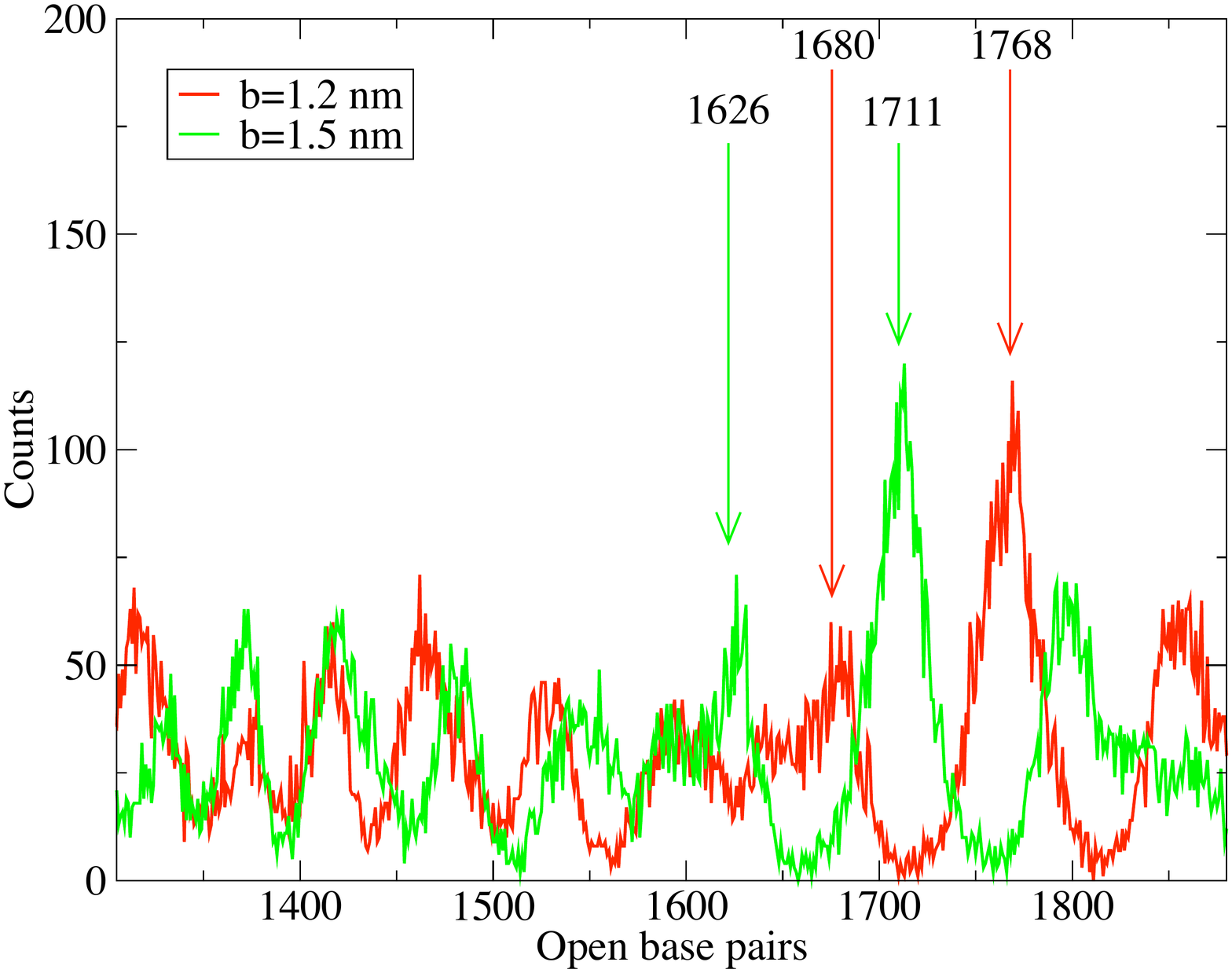}
\caption{Dependence of the histogram on the model parameters. Here we show a region of two histograms of intermediate states that have been calculated using different values of the Kuhn length ($b$) for the ssDNA (see Eq.~(1) in the main text). We highlight two peaks (green and red arrows) of each histogram that correspond to two consecutive intermediate states. The error in the CUR size due to the Kuhn length is less than 4\% (about 3 bps error in a 80 bp-sized CUR).}\label{fig:kuhnhist}
\end{figure}

Here we will discuss what level of uncertainty is introduced in calculating the number of base pairs unzipped through the use of model parameters. The contribution of the handles can be neglected as they behave almost like rigid rods (their contour length is much shorter than the persistence length). We have calculated the same histograms varying the Kuhn length of the ssDNA. When the Kuhn length is modified, the peaks of the histogram are located in a different position. The error introduced might be as large as $\sim$60 bp when the Kuhn length is varied from 1.2 nm to 1.5 nm. However, the difference between the position of two correlative peaks is weakly affected by the Kuhn length ($\sim$3 bp). Fig. \ref{fig:kuhnhist} illustrates these results. Finally, we do not expect important differences in the determination of the CUR sizes by using a slightly different value of the interphosphate distance $d$ because a correction in that value is somehow equivalent to a change in the value of the Kuhn length.

\subsection{Reproducibility of intermediate histograms}
\begin{figure}[!h]
\includegraphics[width=10cm]{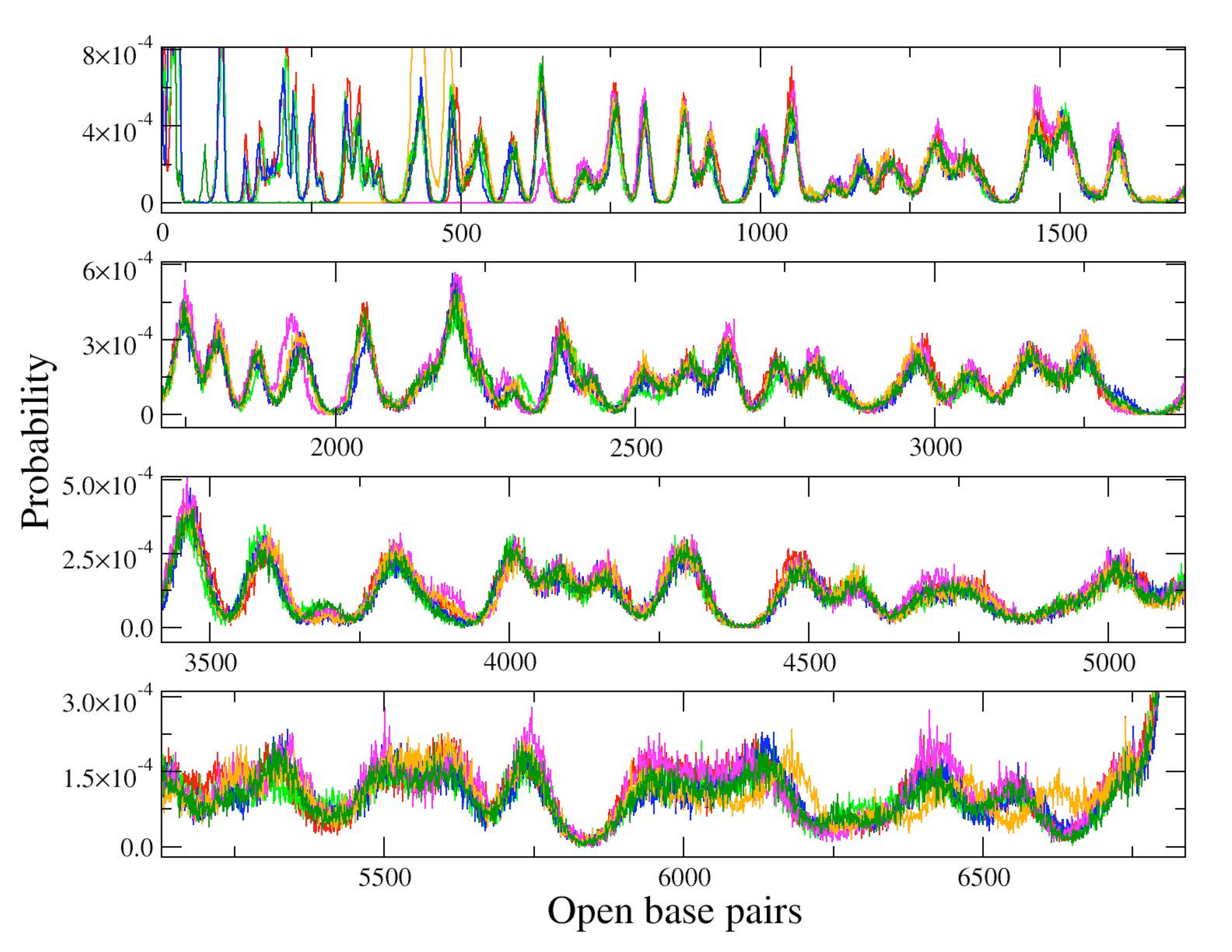}
\caption{Histograms of intermediate states for six different molecules of 6.8 kb. They are depicted in red, green, blue, magenta, orange and dark green. Although the height of each peak is different for the six histograms, the position of the peak is almost the same ($\pm$10~bp). The histograms for the 2.2 kb molecules have similar reproducibility.}\label{fig:histograms6kb}
\end{figure}

Histograms in fig \ref{fig:histograms6kb} show the probability of intermediate states for the 6 different molecules of 6.8 kb (each molecule corresponds to one color). Despite of the fact they look very similar, there are some differences at the beginning (between 0 and 650 bp), mainly due to two reasons: 1) Molecular frying. Some molecules are not capable of completely refold into the DNA duplex and sometimes the first 50-100 bases of the stem remain open. 2) Adhesion between the two beads. The two beads must be very close each other when the DNA is fully zipped because the handles of the molecular construct are very short. Sometimes the beads get stuck and the firsts rips of the unzipping curve cannot be detected. 

\subsection{Error in CUR size distributions}
\begin{figure}[!h]
\includegraphics[width=10cm]{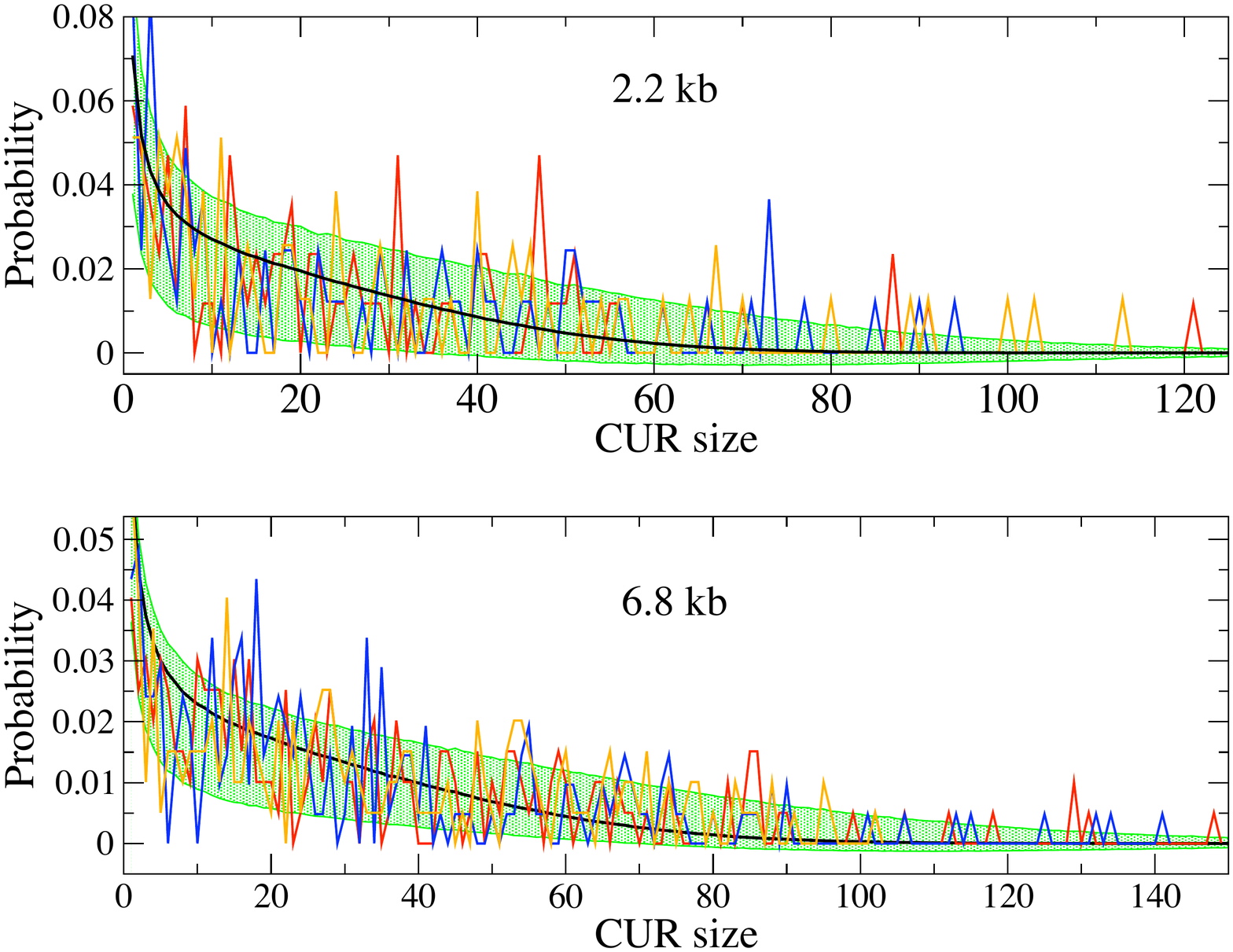}
\caption{CUR size distributions calculated with the toy model. Upper (lower) panel shows the results for a 2252 (6838 kb) sequence. The black curve shows CUR size distribution averaged over $10^4$ realizations. The green region represents the upper and lower limits of the error bars that correspond to the standard deviation of those realizations. Red, blue and orange curves show 3 different realizations. Note the large deviations from the average  histogram due to the finite length of the sequences.}\label{fig:errhist}
\end{figure}

The experimental error of CUR size distributions is negligible as intermediate histograms are fully reproducible among different molecules (see fig.~\ref{fig:histograms6kb}). However we can estimate the error committed in measuring the CUR size distribution among different sequence realizations of the same DNA length. Although that error should vanish for infinitely long sequences (CUR size distributions are self-averaging in the thermodynamic limit) there are large fluctuations for finite length molecules. An experimental measurement of unzipping curves for many DNA sequences is beyond our capabilities. However to estimate that error we can use the the toy model to determine the expected standard deviation of the CUR size distributions (see Fig.~\ref{fig:errhist}).

\subsection{Discrepancies between experimental and theoretical CUR size distributions}
Discrepancies between the experimental results and the mesoscopic model are attributed to two factors: 1) Small CUR are missed due to limited instrumental resolution as described in the paper; 2) Medium and large CUR sizes are prone to large error because less than 10 bp CUR are seldom detected. Indeed, the power law describing the CUR size distributions indicates that the majority of CUR is small sized. However if one small sized CUR is missed then medium or large sized CUR will be overcounted as they should split into smaller pieces whenever they contain a small CUR. It is a difficult math problem to evaluate the final effect of all missed CUR in the resulting CUR size histogram. These two effects concur to modify the shape of the power law for the 6.8kb molecule in Fig.~2d. Yet it is remarkable that the general trend of the experimental data shown in Figs.~2c and 2d follows reasonably well the predictions of the mesoscopic model. This is specially true for the 2.2 kb data shown in Fig.~2b where the non-monotonic oscillations observed in the experimental distribution are captured by the mesoscopic model.

\section{The toy model}

\subsection{Approximated solution to the toy model}

\begin{figure}
\includegraphics[width=10cm]{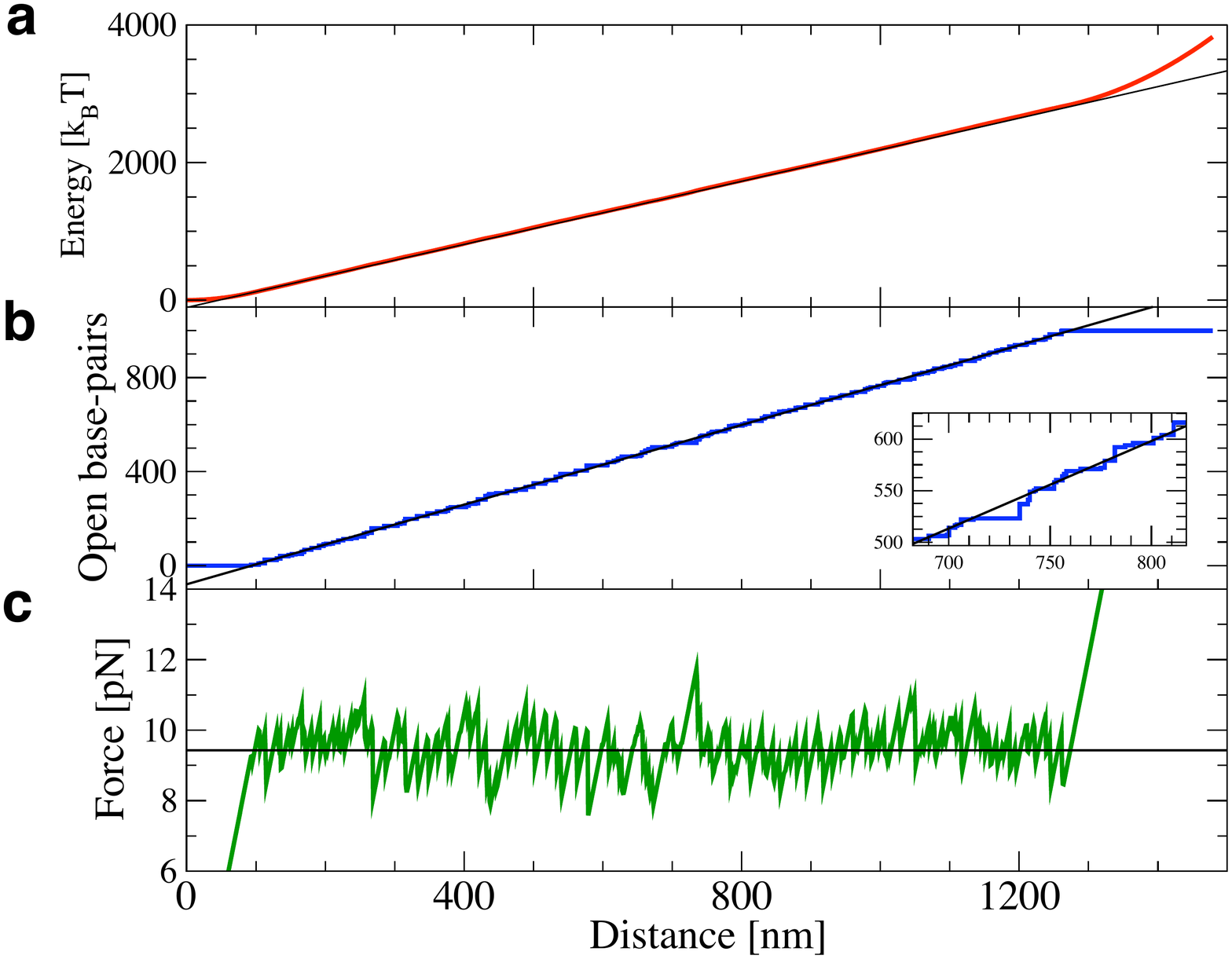}
\caption{Average behavior and one realization of the toy model. Black curves in all the panels show the approximated solution (see main text for the expressions) and colored curves show the solution for one disorder realization. The following parameters have been used: $k$=60~pN/$\mu$m, $d$=0.59~nm, $\mu$=-1.6~kcal/mol, $\sigma$=3.20~kcal/mol.  \textbf{(a)} Energy minimum $E_m(x_{tot})$. \textbf{(b)} Number of open base pairs vs. total distance. Inset shows a detailed view of the stair-shaped character of the curve. \textbf{(c)} FDC.  \label{fig:toymodel}}
\end{figure}

The toy model (Eq.~(4) in the main text) gathers all the relevant features of a DNA unzipping experiment. The most interesting of them are the force rips in the FDC and the discontinuous opening of base pairs (i.e. the CUR). In contrast, equation~(5) in the main text is an approximation that ignores the sequence dependence. The solution to this approximation are smooth expressions that collect the average behavior of the system over an ensemble of sequences (i.e. realizations of the disorder). Figure~\ref{fig:toymodel} shows the approximated solution superimposed on one disorder realization.

\subsection{Size distribution of CUR}

This section shows the results of the simulation of the toy model
introduced in the main text. The distribution of sizes of the CUR
depends on the parameters of the model. The aim of this section
is to characterize such dependency. Starting from the energy contribution of the model (see Eq. 4 in main text for further details), we generate random realizations and obtain the CUR size distribution for each
realization of the disorder. After collecting all the simulated data we obtain the
averaged size distribution of the CUR for the chosen parameters. By
varying the parameters of the model along a wide range we observe how
the shape of the CUR size distribution changes.

In all our simulations we took $d=0.59$ nm and
$\mu=-1.6$ kcal/mol constant, since the distribution of CUR sizes
weakly depends on them. Therefore we only changed $\sigma$ (the
standard deviation of the random distribution of energies) and $k$ (the stiffness of the optical trap). We simulated
sequences of $10^4$ base pairs and we made $10^4$ realizations for each value of $\sigma$ and $k$. 

\subsubsection{Dependence on $\sigma$}

\begin{figure}
\includegraphics[width=9cm]{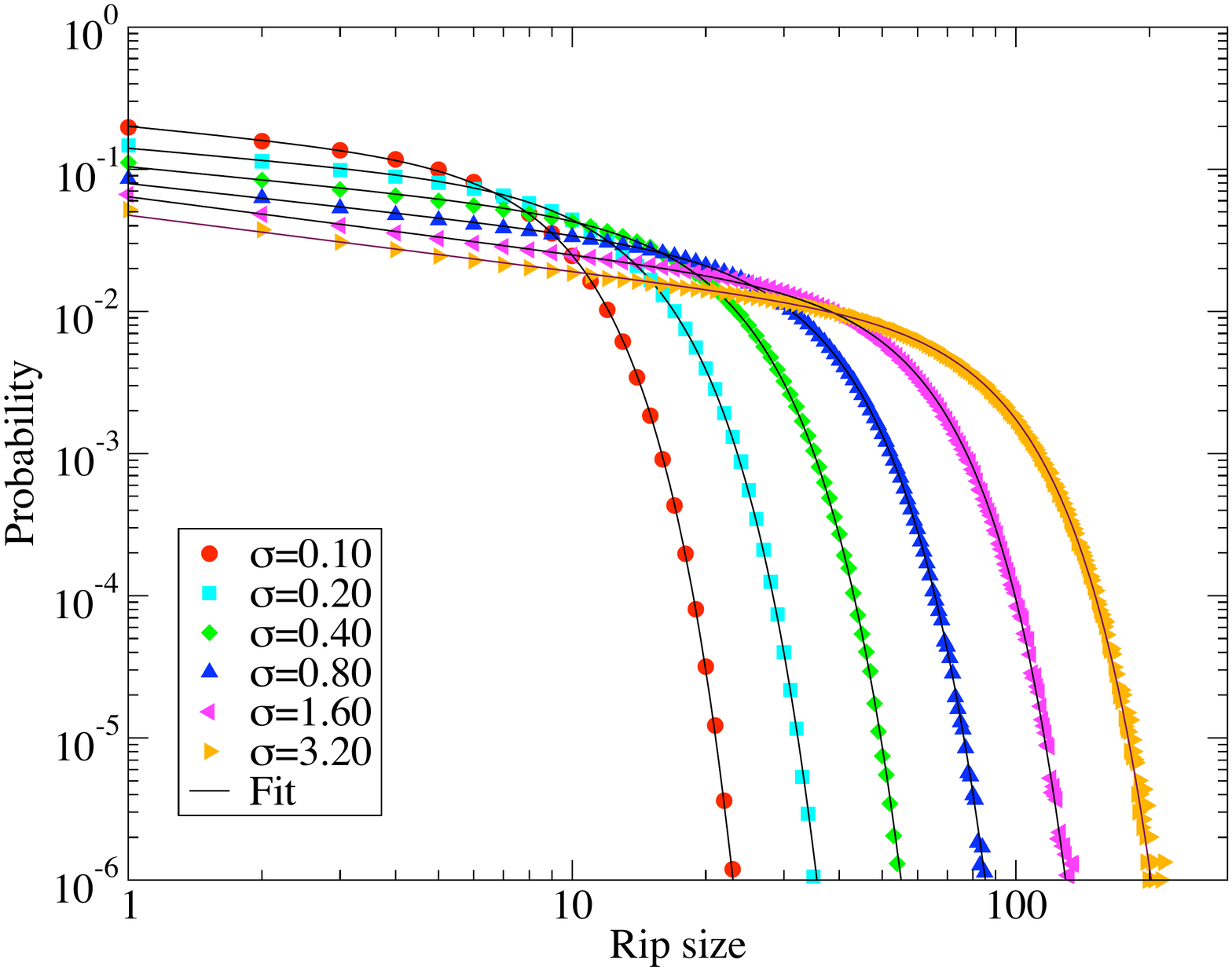}
\caption{CUR size distributions for different values of $\sigma$. The black curves show the fit of Eq. (6) in main text. \label{fig:CURsigma}}
\end{figure}

\begin{figure}
\includegraphics[width=9cm]{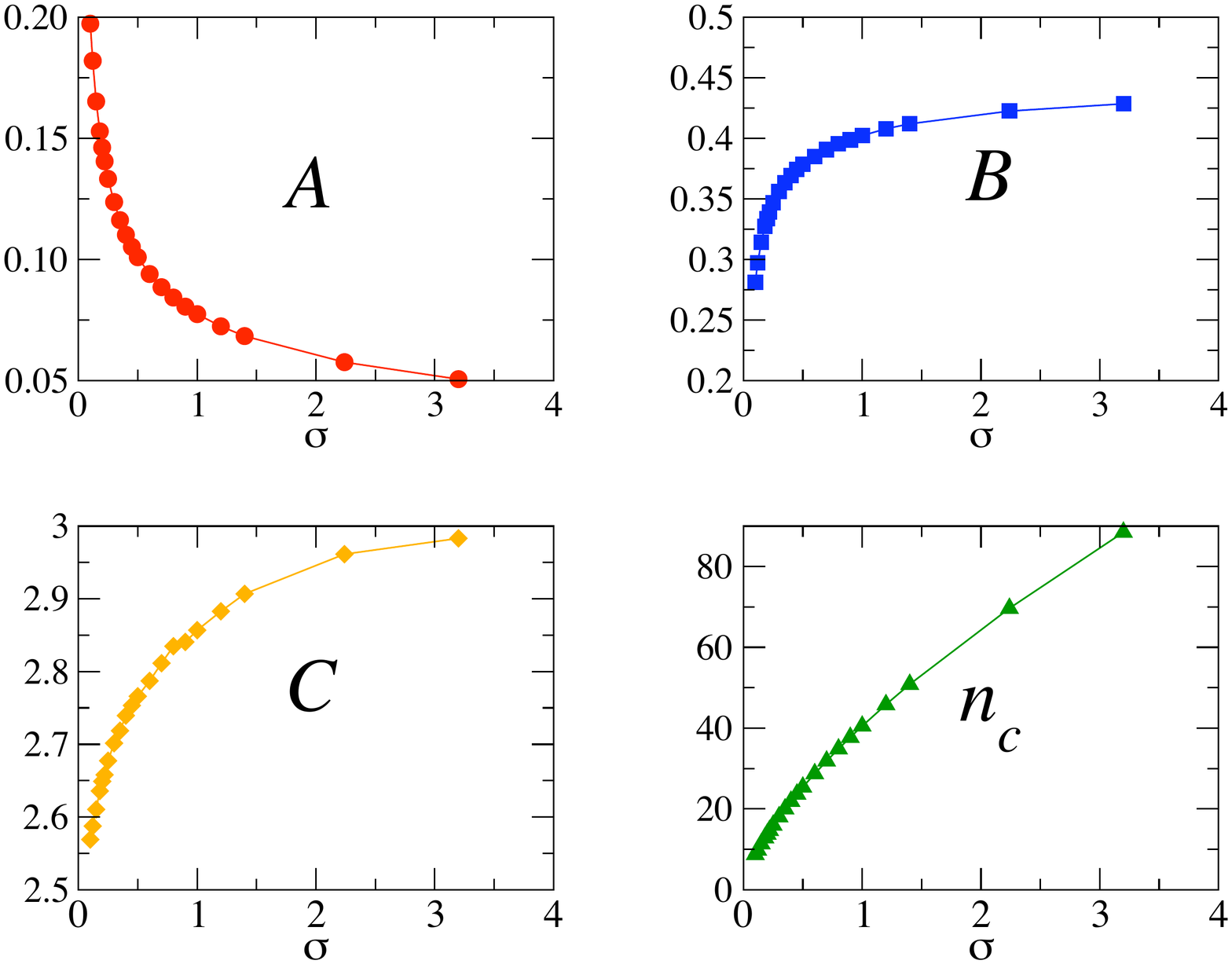}
\caption{Fit parameters plotted versus $\sigma$\label{parameters}.}
\end{figure}

We fixed the trap stiffness at $k$=60 pN/$\mu$m. The distribution of CUR obtained for each value of $\sigma$ is shown in fig~\ref{fig:CURsigma}. The data was fit to Eq.~(6) in main text, where a set of 4 parameters ($A,B,C,n_c$) was obtained for each value of the parameter $\sigma$. Fig.~\ref{parameters} shows the dependence of these parameters with $\sigma$.

\subsubsection{Dependence on $k$}

\begin{figure}
\includegraphics[width=9cm]{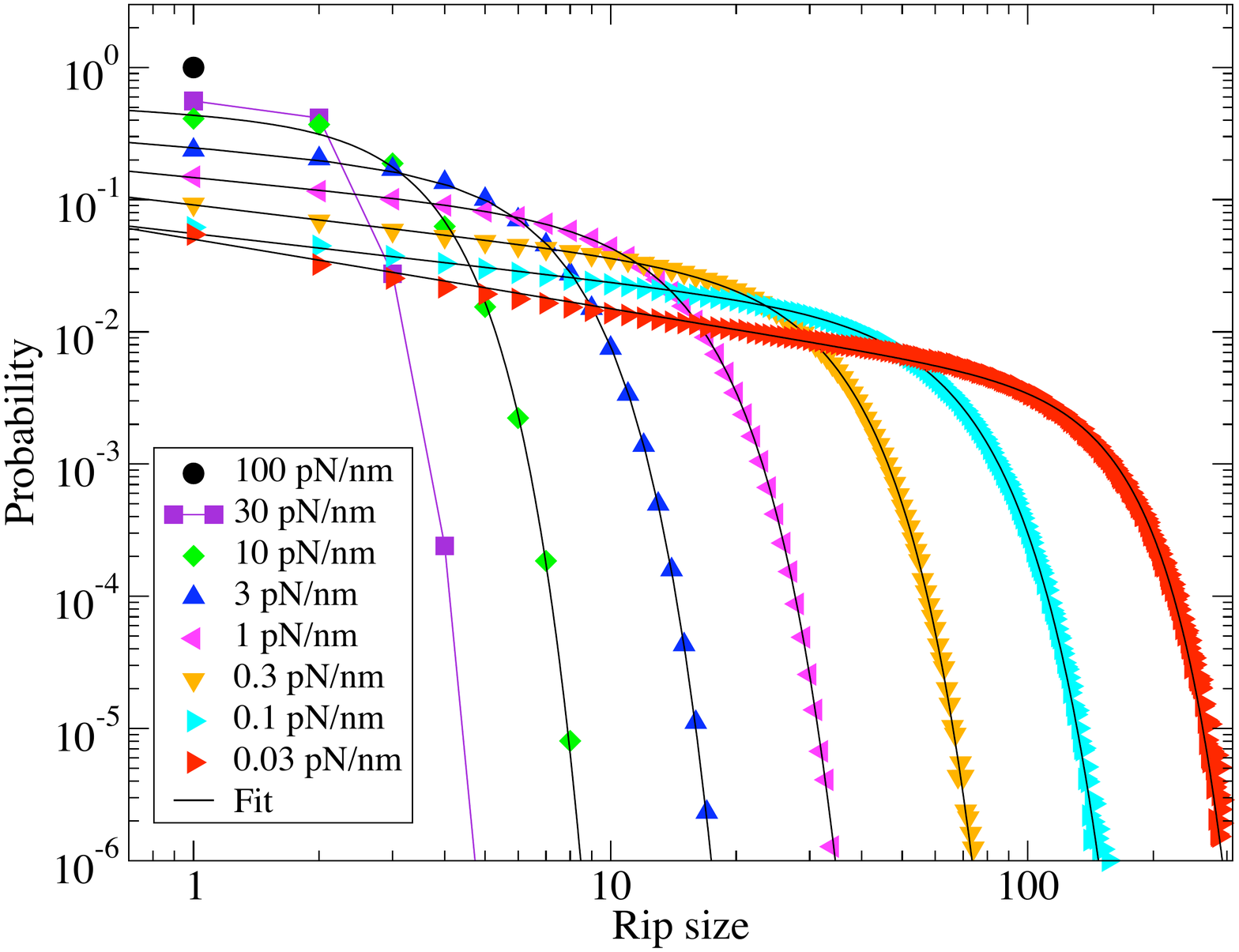}
\caption{CUR size distributions for different values of $k$. The black curves show the fit to Eq. (6) in main text. The tails of the CUR distributions for trap stiffnesses higher than 3 pN/nm need many realizations to be accurately inferred. The distributions are too narrow and the fit of Eq.~(3) does not converge easily. Note that at $k=100$ pN/nm all CUR are one base pair sized. \label{fig:CURk}}
\end{figure}

\begin{figure}
\includegraphics[width=9cm]{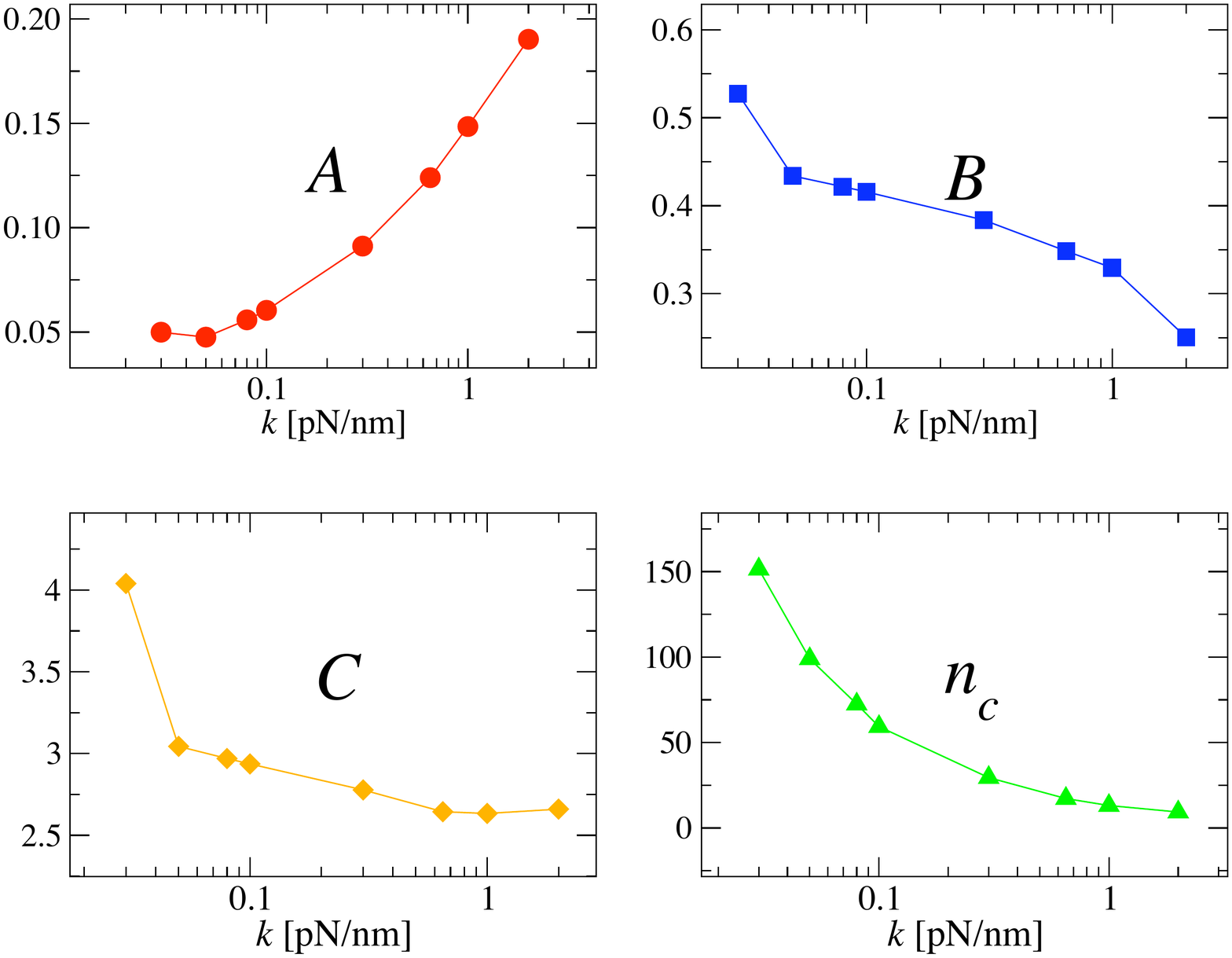}
\caption{Fit parameters plotted versus $k$. The cutoff CUR size ($n_c$) vs. $k$ follows a power law behavior (see Fig.~3d in main text for a log-log plot).\label{fig:parsk}}
\end{figure}

We fixed the amount of disorder at $\sigma$=3.20 kcal/mol. Figure~\ref{fig:CURk} shows the distribution of CUR for some values of $k$ and their fit to Eq.~(6) in main text. Note that in the low $k$ range the CUR size distributions are wide and have good statistics to extract the values for $A,B,C,n_c$. However, for $k>5$ he CUR size distributions are too narrow to be reliably fit to Eq.~(6). Figure~\ref{fig:parsk} shows the dependence of the four parameters ($A,B,C,n_c$) on the trap stiffness.

\section{Stiffness of one nucleotide}

\begin{figure}[!h]
\includegraphics[width=8cm]{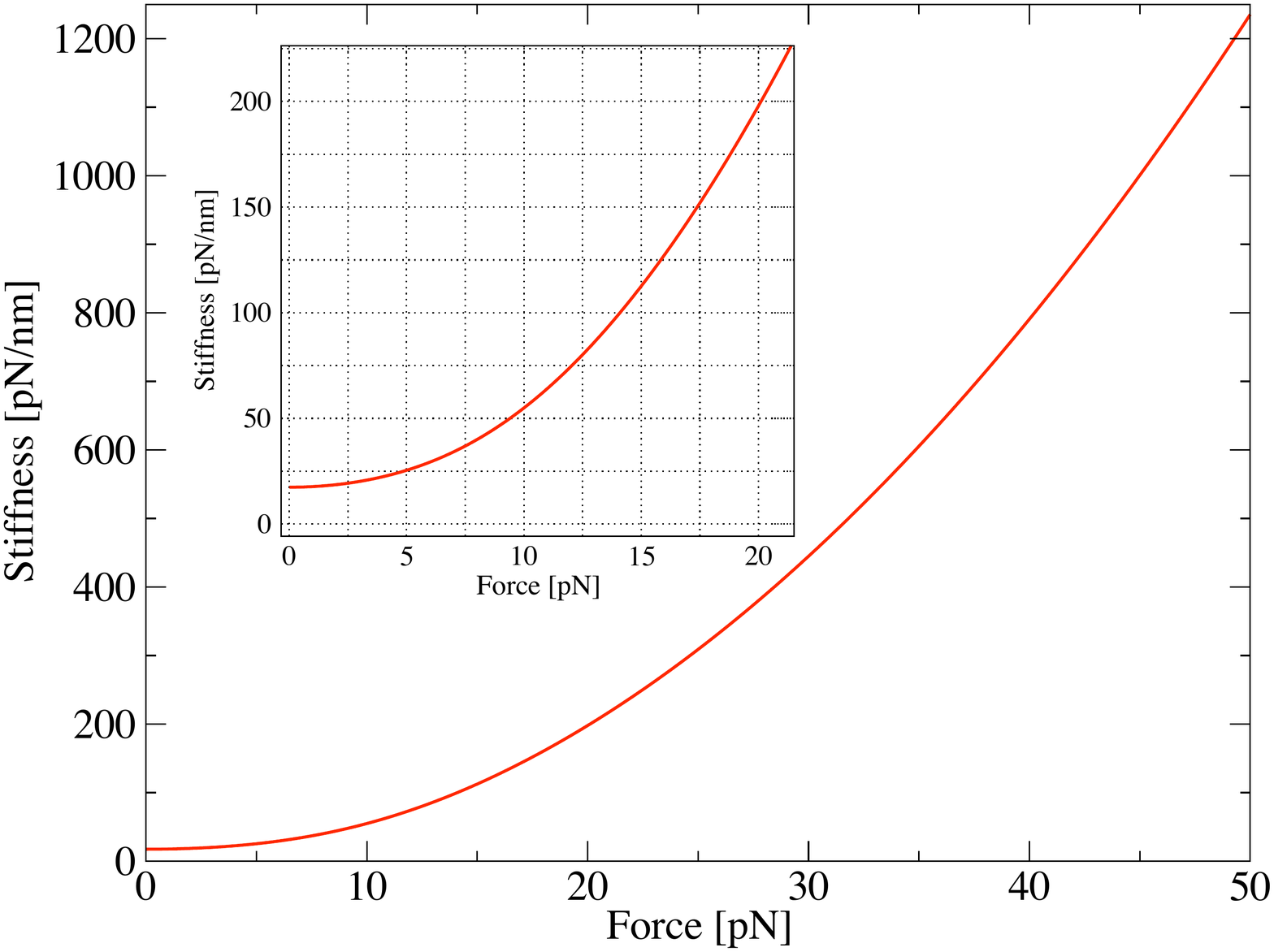}
\caption{Stiffness of one nucleotide vs. force. Inset shows a zoomed section of the curve around the unzipping force.}\label{fig:stiffvsforce}
\end{figure}

Here we calculate the expected stiffness of one nucleotide of ssDNA. The numerical value has been calculated from the elastic response of Freely Jointed Chain (FJC) model for semiflexible polymers, which is given by the following Extension vs. Force curve,
\begin{equation}
x_s(f)=L_0\left(\coth\left(\frac{bf}{k_BT}\right)-\frac{k_BT}{bf}\right)
\end{equation}
where $x_s$ is the extension, $f$ is the force applied at the ends of the polymer, $L_0$ is the contour length, $b$ is the Kuhn length, $k_B$ is the Boltzmann constant and $T$ is the temperature. In the case of a polymer, the contour length ($L_0$) can be written in terms of the number of monomers ($n$) times the length of one monomer ($d$) according to
\begin{equation}
L_0=n\cdot d
\end{equation} 
In the case of a ssDNA molecule, $n$ is the number of bases and $d$ is the interphosphate distance of one nucleotide. The FJC model assumes that the elastic response of the polymer scales with the number of bases. Therefore, the resulting Extension vs. Force expression is a homogeneous function with respect to the number of bases. The stiffness of the polymer at each stretching force is the derivative of the force with respect to the extension $k_s(f)=df/dx_s=(dx_s/df)^{-1}$. For the FJC model, the stiffness is given by the following expression
\begin{equation}
k_s(f)=\left[ n\cdot d\left(-\frac{b}{k_BT}\textrm{cosech}^2\left(\frac{bf}{k_BT}\right)+\frac{k_BT}{bf^2}\right)\right]^{-1}
\end{equation}
Using the parameters from section \ref{sec:parameters} ($b=1.2$~nm, $d=0.59$~nm) for one nucleotide ($n=1$) we get a stiffness of $k_s=113$~pN/nm at $f=15$~pN and $k_s=127$~pN/nm at $f=16$~pN (see Fig.~\ref{fig:stiffvsforce}).

\section{Protein-DNA interaction}

In the cell, the function of helicases is to unzip DNA during the replication process. Although their mechano-chemistry is not clear \cite{Lionnet:2006aa} we interpret that helicases pull directly on the ssDNA. In this simplified view of the process, we visualize the helicase as a clamp that slides along one strand of the DNA and applies local force at the unzipping fork. In a more general scheme, the helicase applies force on the DNA by means of an effective stiffness $k^{-1}\sub{eff}=k^{-1}_h+k^{-1}_s$, where $k_h$ is the stiffness of the helicase and $k_s$ is the stiffness of one base of ssDNA. From the conclusions of our work, we know that $k_s$ is high enough to locally unzip DNA one bp at a time. Therefore, the unzipping process will be one bp at a time as long as $k_h$ is higher than $k_s$. Indeed, when the helicase pulls directly on DNA the stiffness of the helicase can be assumed to be very large (proteins are indeed very rigid objects) compared to the stiffness of a single base pair ($k^{-1}_h\ll k^{-1}_s$) and the effective stiffness between the helicase and the DNA is approximately equal to the stiffness of ssDNA ($k\sub{eff} \sim k_s$).

The previous explanation can be extended to proteins that interact with DNA. If a protein increases the stiffness of one bp of ssDNA, the local unzipping still could be done one bp at a time. On the other hand, if a protein decreases the ssDNA stiffness below the boundary of $k_s\sim 100$~pN/nm the local unzipping would show CUR of sizes larger than one bp. As far as we know, there is no protein with high compliance bound to the ssDNA between the helicase and the unzipping fork when the replication complex (helicase, polymerase, etc.) is set. However the full scenario of what might happen for different biological models under varied conditions remains to be seen.

\section{CUR and genes}

\begin{figure}[!h]
\includegraphics[width=13cm]{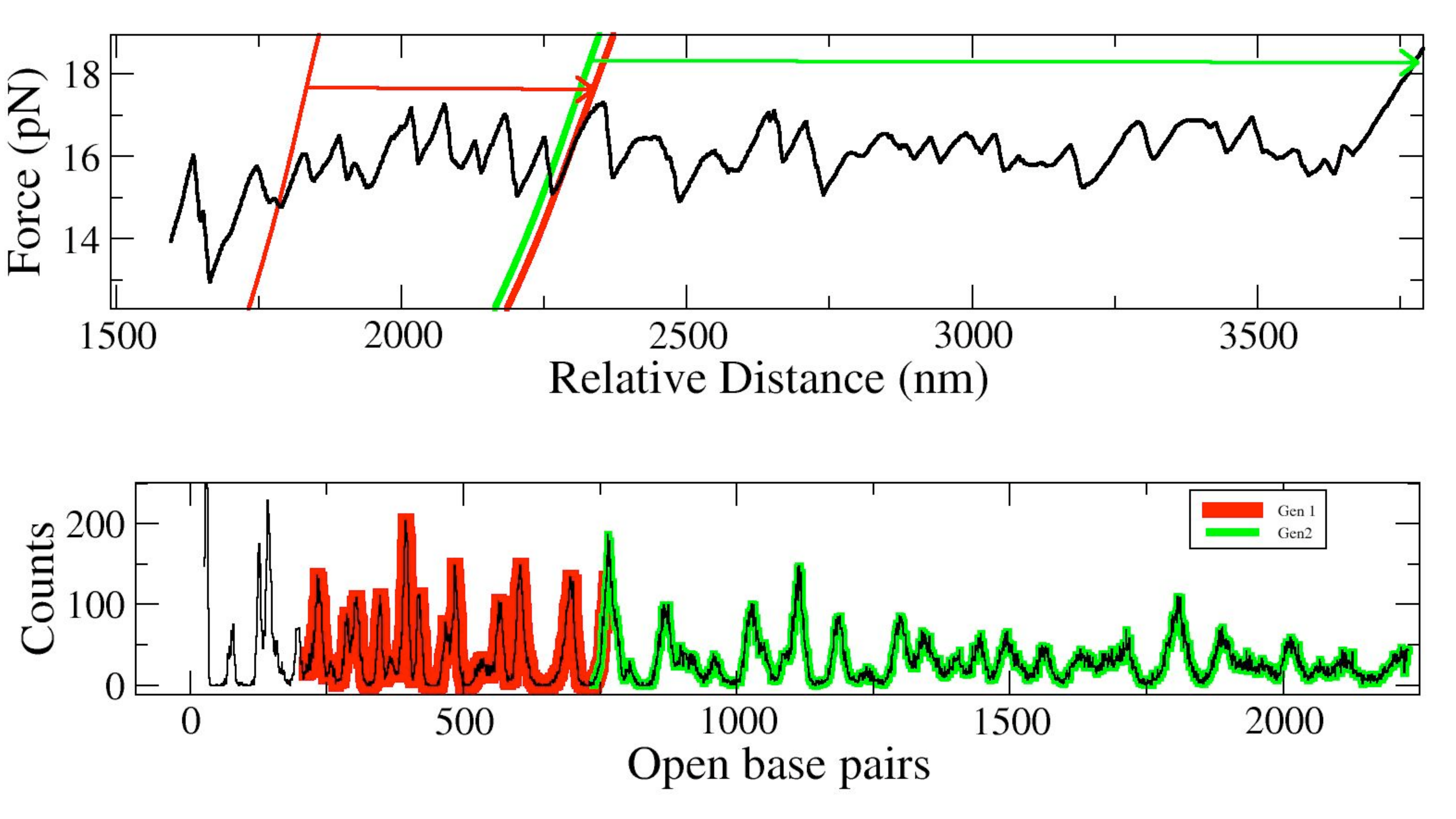}
\caption{Localization of genes for the 2.2 kb sequence. \textbf{(a)} Start and end points of each gene. The arrow shows the transcription direction. Each gene is depicted in a different color. \textbf{(b)} Position of the genes along the histogram of number of unzipped base pairs.\label{gens2kb}.}
\end{figure}

\begin{figure}
\includegraphics[width=13cm]{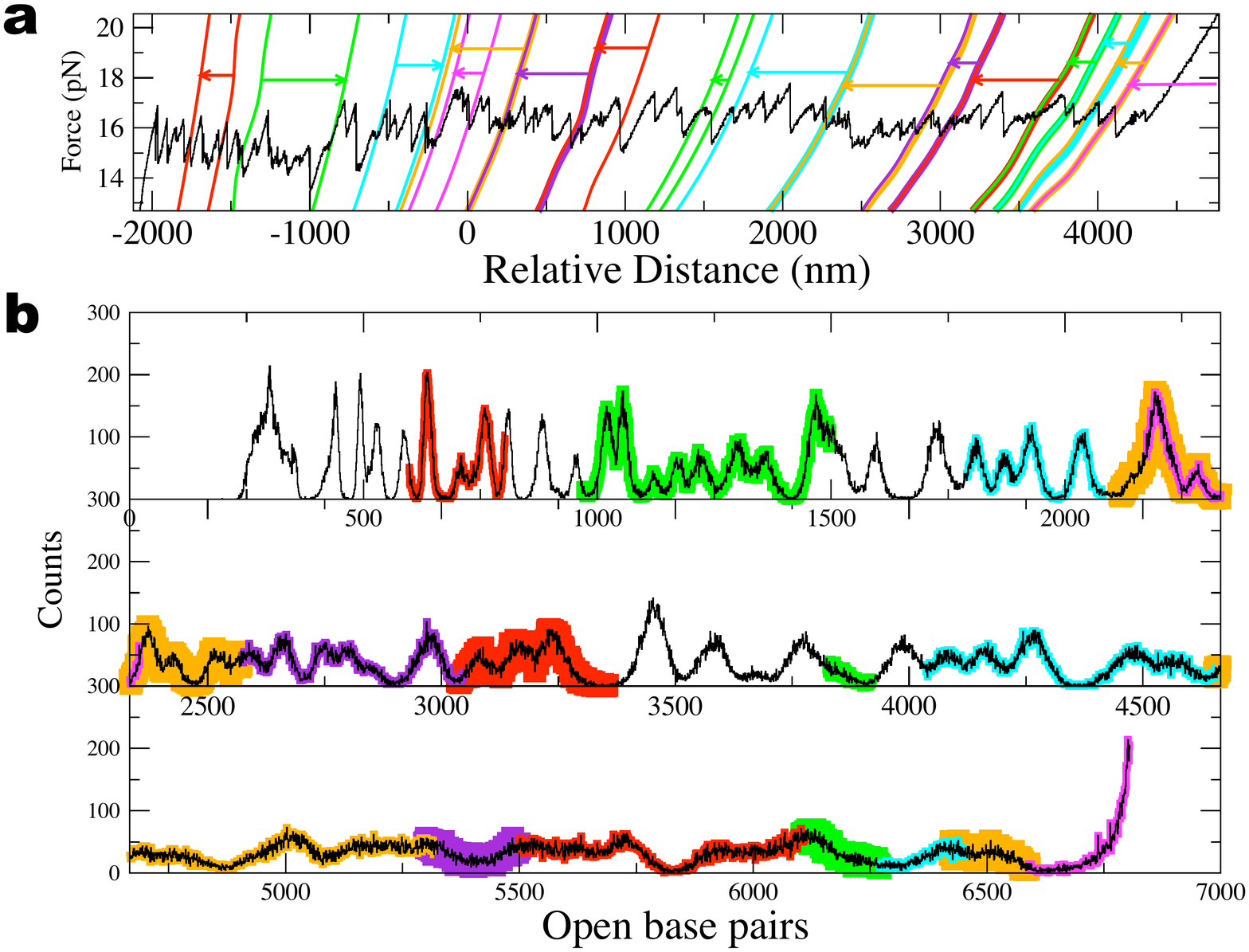}
\caption{Localization of genes for the 6.8 kb sequence. Same color code as in Fig. \ref{gens2kb}\label{gens6kb}.}
\end{figure}

As an extra information to the reader, here we show the position of the genes  that are localized in the two fragments of $\lambda$-DNA used in this work. The 2.2~kb fragment contains partially one gene and one
complete gene. Upper panel in Fig.~\ref{gens2kb} shows the localization of the
genes along the Force Distance Curve. Lower panel in Fig.~\ref{gens2kb} shows the
position of the genes superimposed on the histogram of number of
unzipped base pairs. On the other hand, the 6.8 kb fragment contains
partially one gene and 15 complete genes. Figure \ref{gens6kb} shows
the location of these genes superimposed on the histogram of number
of unzipped base pairs. It can be clearly observed that the lengths of
most of these genes span over several rips. 

We should not expect correlations between the CUR and the genes because the CUR depend on the trap stiffness used in the experimental setup. In other words, a different trap stiffness produces a different distribution of CUR on the same DNA molecule.

\newpage

\bibliographystyle{unsrt}